\documentclass[aps,prl,twocolumn,amsmath,amssymb,%
longbibliography]{revtex4-2}
\usepackage{graphicx}
\usepackage{epsfig}
\usepackage{dcolumn}
\usepackage{bm}
\usepackage{bbm}
\usepackage[normalem]{ulem}
\usepackage{color}
\usepackage{float}
\usepackage[colorlinks=true,linkcolor=blue,citecolor=blue,urlcolor=blue,breaklinks=true]{hyperref}%
\usepackage{verbatim}
\usepackage{subfigure}
\usepackage[ruled,vlined]{algorithm2e}
\usepackage{tikz}
\usetikzlibrary{calc}
\usetikzlibrary{shapes}
\usetikzlibrary{arrows.meta}
\tikzset{
  fractal line/.style args={#1 and #2}{%
    to path={
      let
      \p1=(\tikztostart), %
      \p2=(\tikztotarget), %
      \n1={veclen(\x1-\x2,\y1-\y2)}, %
      \p3=($(\p1)!.5!(\p2)$), %
      \p4=(rand*#1*\n1,rand*#1*\n1), %
      \p5=(\x3+\x4,\y3+\y4) %
      in \pgfextra{
        \pgfmathtruncatemacro\mytest{(\n1<#2)?1:0}
        \ifnum\mytest=1 %
        \tikzset{fractal line/.style args={#1 and #2}{line to}}
        \fi
      } to[fractal line=#1 and #2] (\p5) to[fractal line=#1 and #2] (\p2)
    },
  },
}

\usepackage{braket}

\def\trace#1{{\operatorname{Tr}}\left[#1\right]}
\def\be{\begin{equation}}       \def\ee{\end{equation}}
\def\bea{\begin{eqnarray}}      \def\eea{\end{eqnarray}}
\def\ba{\begin{array}}
\def\ea{\end{array}}
\def\bnum{\begin{enumerate} }
\def\enum{\end{enumerate}}

\def\=>{\Rightarrow}
\def\>{\rightarrow}

\def\eye2{Fathbb{I}}

\usepackage{ listings }

\renewcommand{\>}{\rangle}

\newcommand{\p}{\partial}

\newcommand{\pfaffian}{\text{Pf}}
\newcommand{\D}{\text{D}}
\newcommand{\ima}{\mathrm{i}}

\newcommand{\eq}[2]{
	\begin{equation}
	#1 \label{#2}
	\end{equation}
}

\newcommand{\vect}[1]{\boldsymbol{#1}}

\usepackage{listings}
\usepackage{color}
\definecolor{lightgray}{gray}{1}

\begin{document}

\title{Pfaffian quantum Monte Carlo: solution to Majorana sign ambiguity and applications}
\author{Ze-Yao Han}
\altaffiliation{The two authors contributed equally to this work.} 
\affiliation{Institute for Advanced Study, Tsinghua University, Beijing 100084, China}
\author{Zhou-Quan Wan}
\altaffiliation{The two authors contributed equally to this work.} 
\affiliation{Institute for Advanced Study, Tsinghua University, Beijing 100084, China}
\author{Hong Yao}
\email{yaohong@tsinghua.edu.cn}
\affiliation{Institute for Advanced Study, Tsinghua University, Beijing 100084, China}

\begin{abstract}
Determinant quantum Monte Carlo (DQMC), formulated in complex-fermion representation, has played a key role in studying strongly-correlated fermion systems. However, its applicability is limited due to the requirement of particle-number conservation %
after Hubbard-Stratonovich transformation. %
In going beyond the conventional DQMC, one encouraging development occurred when Majorana fermions were introduced for QMC \cite{Li2015PRB,Li2016PRL}. 
But in previous Majorana-based QMC, Boltzmann weight is determined often with a sign ambiguity. %
Here we successfully resolved this ambiguity by deriving a {\it closed-form} Pfaffian formula for the weight, enabling efficient calculation of the weight with its sign in polynomial time.
We call it ``Pfaffian quantum Monte Carlo'' (PfQMC), which can be applied to generic interacting fermion models. We have successfully employed PfQMC to explore how robust Majorana edge modes in Kitaev chain are against strong interactions. 
By offering greater flexibility, PfQMC can potentially enhance existing sign-mitigating and approximation methods and help address challenging issues such as the ground-state properties of the doped Hubbard model.

\end{abstract}

\date{\today}
\maketitle

{\it Introduction.---} Quantum Monte Carlo (QMC) methods \cite{blankenbecler1981monte,scalapino1981monte, sandvik1991, sse1999loop, DMC2001RMP,Hirsch1985prb,white1989numerical} are often of vital importance in the studies of quantum many-body systems, especially those with strong interactions. Among these, the determinant quantum Monte Carlo (DQMC), pioneered by Blankenbecler, Scalapino, and Sugar \cite{blankenbecler1981monte,scalapino1981monte} stands out as one of the most successful approaches in dealing with fermionic systems. %
DQMC can conduct intrinsically {\it unbiased} simulations and yield numerically-exact results. %
While DQMC is quite versatile, it implicitly adhere to a particle-number-conserving symmetry.
This is because, in DQMC, an interacting problem is always transformed to a summation or integration over non-interacting problems through Hubbard-Stratonovich (HS) transformation \cite{hubbard1959calculation,stratonovich1957method,hirsch1983discrete}.
Only when there is a conserved particle number can the trace of non-interacting problems be straightforwardly evaluated as a determinant, thereby enabling subsequent Monte Carlo simulations. %
This limitation prevents us from numerous possibilities, especially the treatment of many-body Hamiltonians that contain explicit pairing fields (unless an effective U(1) symmetry can be restored under certain transformations \cite{Otsuka2020prb,Xu2021prl}). %

In going beyond the complex-fermion approach with particle-number conservation in conventional DQMC, one encouraging development occurred when Majorana fermions were introduced such that the pairing terms and particle-number preserving terms are treated on an equal footing \cite{Li2015PRB,Li2016PRL}. %
This Majorana quantum Monte Carlo (MQMC) method has greatly extended the realm of sign-problem-free simulations \cite{li2015fermion, Hesselmann2016prb,Assaad2016prx,Li2017NC,Li2017prb,Li2017prl,Xu2023prr,Li2024prl}, and has successfully led to a sufficient condition which could be used to classify all sign-problem-free models known by now \cite{Li2016PRL, wei_majorana_2016,wei2018semigroup, ZXLiQMCreview}. 
However, for models which cannot be proved to be sign-free in MQMC, a sign ambiguity arises when calculating the Monte Carlo weight \cite{Li2015PRB}, preventing it from handling those sign-problematic models. 
For instance, the current method cannot simulate the one-dimensional (interacting) Kitaev chain \cite{Kitaev_2001} or the two-dimensional spinless $p+ip$ (interacting) topological superconductors \cite{Read2000prb}.

Sign-problematic simulations are indeed challenging. Nevertheless, recent evidence suggests that they are not only feasible for studying interesting physics (see, e.g. Refs. \cite{huang2017stripe, huang2019strange_metal, sato2021prb, wang2023wiedemann}), but, in certain cases, they can be employed as advantages. 
For example, it has been demonstrated that the prerequisite of being sign-problem-free is overly restrictive and inapplicable to many phases of matter \cite{grossman2023robust,smith2020intrinsic,golan2020intrinsic};
even the sign problem itself has been postulated as a potential indicator of the quantum phase transitions within the underlying system \cite{mondaini2022quantum,mondaini2023universality,Mou2022,li2022asymptotic,Yi2024}.
Despite the general interest in investigating the sign problem itself, many efforts have been devoted to alleviate it, for instance, the sign-mitigating methods \cite{Tremblay1992,wan2022prb,easing2020,Vaezi2021prl,Levy2021} and the constraint-path approximation \cite{zhang1997constrained,stripe2017,yuanyao2019prb,xu2024coexistence}. These approaches often require flexibility in choosing the HS transformation schemes or the trial wave-functions. The extension to Majorana fermions could provide further opportunities to investigate the nature of the sign problem and improve current methods with greater flexibility. It is thus highly desired to resolve this Majorana sign ambiguity and develop a general framework.

In this paper, we elucidate the origin of the Majorana sign ambiguity and present a unified formalism capable of resolving it completely. Building on this formalism, we propose the Pfaffian quantum Monte Carlo (PfQMC) method as a natural and significant extension of DQMC, analogous to how the Pfaffian serves as a more generalized concept compared to the determinant in linear algebra.
As will soon become evident, PfQMC not only goes beyond the capabilities of DQMC but also retains all its efficiency-enhancing features.
We further utilize PfQMC to study the phase transition in the interacting Kitaev chain and examine the behavior of the Majorana-resolved sign in the spinless $t$-$V$ model, as concrete examples of applying PfQMC.

{\it QMC and Majorana sign ambiguity.}---In DQMC, after the Trotter decomposition and HS transformation, the partition function of a generic fermionic many-body system is mapped into a summation of Boltzmann weights over different auxiliary-field configurations $\{\phi_n\}$,
\eq{
\begin{split}
\trace{ e^{-\beta \hat{H}} } 
&= \sum_{\{\phi_n\}} \trace{ \prod^{L_{\tau}}_{n=1} e^{-\hat{h}(n)} } =\sum_{\{\phi_n\}} w({\{\phi_n\}}).\\
\end{split}
}{traceToDet}
Here the partition function is trotterized onto $L_{\tau}$ imaginary time slices, and at each time slice there is a quadratic fermion Hamiltonian $\hat{h}(n) =\sum_{ij} {c}_i^\dagger h_{ij}(n) {c}_j$ that implicitly depends on the auxiliary field configuration $\{\phi_n\}$. The trace can be carried out as $w(\{\phi_n\})\equiv\det\left[\mathbb{I} + \prod^{L_{\tau}}_{n=1} e^{-h(n)} \right]$ which is the weight associated with each $\{\phi_n\}$, or from field theory perspective, the weight of each path in a path integral. Notably, in this formalism particle-number conservation is required to obtain a determinant after tracing out fermions.  %
The fact that this weight can be carried out \textit{analytically} lies in the heart of all DQMC-related algorithms \cite{blankenbecler1981monte, Hirsch1985prb,white1989numerical,zhang1997constrained}.

When Majorana fermions were introduced for QMC \cite{Li2015PRB}, pairing creation and annihilation operators are allowed naturally in the HS-decomposed quadratic Hamiltonians. In terms of Majorana fermions, the quadratic Hamiltonians $\hat h$ can now be re-written in a most general form:
\begin{equation}
    \hat{h}(n) = \frac{1}{4} \bm{\gamma}^T h(n) \bm{\gamma} = \frac{1}{4} \sum_{i, j = 1}^{2N} \gamma_i [h(n)]_{ij} \gamma_j,
\end{equation}
where $2N$ Majorana operators $\bm\gamma^T = (\gamma_1, \cdots, \gamma_{2N})$ represent $N$ complex-fermion degrees of freedom. We adopt the convention that a pair of creation/annihilation operators is transformed to two Majorana operators via $c = \frac{1}{2} (\gamma_1 + \ima \gamma_2)$ and $c^\dagger = \frac{1}{2} (\gamma_1 - \ima \gamma_2)$, hence the anti-commutation relation is given by $\{\gamma_i, \gamma_j\} = 2\delta_{ij}$. Without loss of generality, the $2N \times 2N$ matrix $h (n)$ can then be assumed skew-symmetric, $h^T = -h$. 

As in conventional DQMC formulation, a closed-form expression of the statistical weight, defined also by $w(\{\phi_n \}) = \trace{\prod_n e^{- \hat{h}(n)}}$, shall serve as a vital ingredient for us to proceed. However, previous attempts have only managed to determine this weight up to an ambiguous sign, namely 
\begin{equation}\label{ambiguity}
  w(\{\phi_n \}) = \pm \left\{\det\left(\mathbb{I} + \prod_n e^{- h(n)}\right)\right\}^{\frac{1}{2}},  
\end{equation}
we call this the Majorana sign ambiguity in MQMC. Although 
special instances of this ambiguity have already been addressed \cite{robledo_sign_2009, shi2017many}, it was not finally resolved until our present work.

\newcommand\drawcrak[3][fill=black]{
  \path[#1] (#2) to[fractal line=.04 and 3mm] (#3) to[fractal line=.03 and 4mm] (#2);
}

\begin{figure}[t!]
    \begin{tikzpicture}
        \pgfmathsetseed{114512};
        \drawcrak[fill=black]{{-3, 1}}{{4, 2}};
        \pgfmathsetseed{114515};
        \drawcrak[fill=black]{{0, 2.5}}{{-2, -1.2}};
        \pgfmathsetseed{114514};
        \drawcrak[fill=black]{{3, 3}}{{3.5, -1}};
        
        \node[label=right:{$(\eta_1, B_1 )$}] (element_1) at (0.7, 2.2) {};
        \fill (element_1) circle (2pt);
        \node[label=right:{$(\eta_3, B_3 )$}] (element_3) at (1.7, -0.6) {};
        \fill (element_3) circle (2pt);
        \draw [-{Latex[length=2.5mm]}, line width=.8mm] (element_1) to [bend right=30] node [below, sloped] (TextNode1) {$(\eta_2, B_2)$} (element_3);
    \end{tikzpicture}
    \caption{Schematic illustration of our construction. The figure represents the group manifold of $\text{Spin}(2N, \mathbb{C})$. Every element with non-vanishing trace can be uniquely identified by a pair $(\eta, B)$ satisfying $2^{2N} \eta^2 = \det(I + B)$. Those with vanishing weights are visualized as ``cracks'' on the manifold, which divide the manifold into sectors with $2^{N} \eta = \pm  \sqrt{\det(I+B)}$. Consider the real-valued weight scenario, a product of two finite elements in the Spin group is able to ``jump over'' the cracks , as exemplified by the thick arrow, where product of two FGOs represented by $(\eta_1, B_1)$ and $(\eta_2, B_2)$ results in an FGO $(\eta_3, B_3)$. The Monte Carlo sampling process, can never ``step on'' such cracks, rendering our construction particularly suitable for QMC simulations.}
    \label{fig:PfQMC1}
\end{figure}
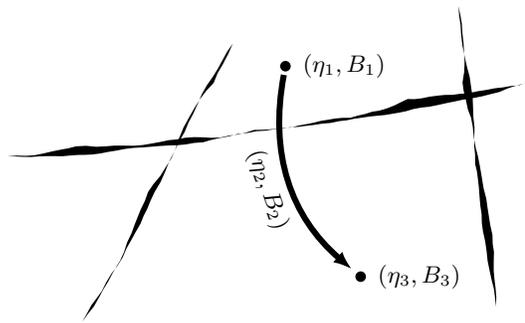

{\it Solution to Majorana sign ambiguity.}---Let us first take a closer look into the origin of this sign ambiguity. Hereafter, operators like $e^{-\frac{1}{4}\bm \gamma^T h \bm \gamma}$ shall be termed fermionic Gaussian operators (FGOs). 
In conventional DQMC algorithm, FGOs $e^{-c^\dagger h c}$ with particle-number conservation can be faithfully represented by invertible matrices $e^{-h}$ because they are nothing but different realizations of the $\text{GL}(N, \mathbb{C})$ group and can be related by a group isomorphism. 
This isomorphism is what lends efficiency and elegance to the DQMC method.
However, for our more general setting here, a $\mathbb{Z}_2$ information is lost if $e^{-h}$ alone is used to represent general FGOs $e^{-\frac{1}{4}\bm \gamma^T h \bm \gamma}$ due to the following group theory observations:
(i) The whole set of FGOs constitute the $\text{Spin}(2N, \mathbb{C})$ group, generated by the Majorana bilinears $\frac{1}{4} \bm{\gamma}^T h \bm{\gamma}$ \cite{ma2007group}. 
(ii) The skew-symmetric matrices $h$ generate the special orthogonal group $\text{SO}(2N, \mathbb{C})$. The Heisenberg equation of motion, $e^{\frac{1}{4} \bm \gamma^T h \bm \gamma } \gamma_i e^{-\frac{1}{4} \bm \gamma^T h \bm \gamma} = \sum_j \left[ e^{-h} \right]_{ij} \gamma_j $, naturally defines a 2-to-1 homomorphism from $\text{Spin}(2N, \mathbb{C})$ to $\text{SO}(2N, \mathbb{C})$. 
(iii) $\text{Spin}(2N, \mathbb{C})$ is a double cover of $\text{SO}(2N, \mathbb{C})$, i.e. $\text{Spin}(2n, \mathbb{C}) / \text{SO}(2n, \mathbb{C}) = \mathbb{Z}_2$. This $\mathbb{Z}_2$ is the source of the sign ambiguity in representing FGOs by special orthogonal matrices \cite{klich_note_2014}.

A naïve attempt to recover this missing $\mathbb{Z}_2$ might involve ``representing'' elements of the Spin group by a pair $(\xi, B)$, where $\xi \in \mathbb{Z}_2$ and $B \in \text{SO}(2n, \mathbb{C})$. 
As will soon become evident, such representation can exist almost everywhere, through the use of the \textit{Grassmann representation} of FGOs \cite{bravyi_lagrangian_2004}.
Below, we present our main results (see the Supplementary Material for detailed derivation): %

In the Grassmann representation, an FGO $\mathcal{O}$ can be represented by a pair $(\eta, B)$, where $\eta$ is a c-number such that $\text{Tr}[\mathcal{O}] = 2^N \eta$ and $B \in \text{SO}(2n, \mathbb{C})$. 
For a specific FGO $\mathcal{O}=e^{-\frac{1}{4} \bm{\gamma}^T h \bm{\gamma}}$, $B = e^{-h}$, and $\eta$ can be straightforwardly calculated as:
\begin{equation}
    \eta(h) = (-1)^N \pfaffian{\begin{pmatrix}
        \sqrt{2} \sinh(\frac{h}{4}) & -I \\
        I & \sqrt{2} \sinh(\frac{h}{4})
    \end{pmatrix}}.
    \label{eq:GRepEtaVal}
\end{equation}

For two FGOs ${\mathcal{O}}_1, {\mathcal{O}}_2$ with representations $(\eta_1,B_1)$ and $(\eta_2, B_2)$, the product $ \mathcal{O}_3 = \mathcal{O}_1  \mathcal{O}_2$ is also an FGO, and its representation can be derived as follows:

\begin{equation}
    \eta_3 = (-1)^N \eta_1 \eta_2 \pfaffian \begin{pmatrix}
        G_1 & -I \\ I & G_2 \end{pmatrix}, \, B_3=B_1B_2,
    \label{eq:productRuleMain}
\end{equation}
where $G_1$ and $G_2$ are Green functions associated with $\mathcal{O}_1,\mathcal{O}_2$,  defined by $G(\mathcal{O})_{ij}\equiv \trace{\mathcal{O} \gamma_i \gamma_j}/\trace{\mathcal{O}} $ whenever $i \neq j$ and $G_{ii} = 0$. The Green function is thus a skew-symmetric matrix and is related to the matrix $B$ as $G = 2 (I + B)^{-1} - I$. One can verify that $(\eta, B)$ satisfies the constraint $2^{2N} \eta^2 = \det (I+ B)$. In this sense the missing $\mathbb{Z}_2$ is retrieved. Eq.~\ref{eq:GRepEtaVal} and Eq.~\ref{eq:productRuleMain} have enabled us to 
carry out the aforementioned Monte Carlo weight $w(\{\phi_n\})$ in an iterative way. Given that evaluating a single weight requires computing $L_{\tau}$ Pfaffians, the time complexity for this procedure is $O(N^3 L_{\tau})$ \cite{Wimmer_2012}.

\begin{figure*}[t!]
    \includegraphics*[width=0.9\linewidth]{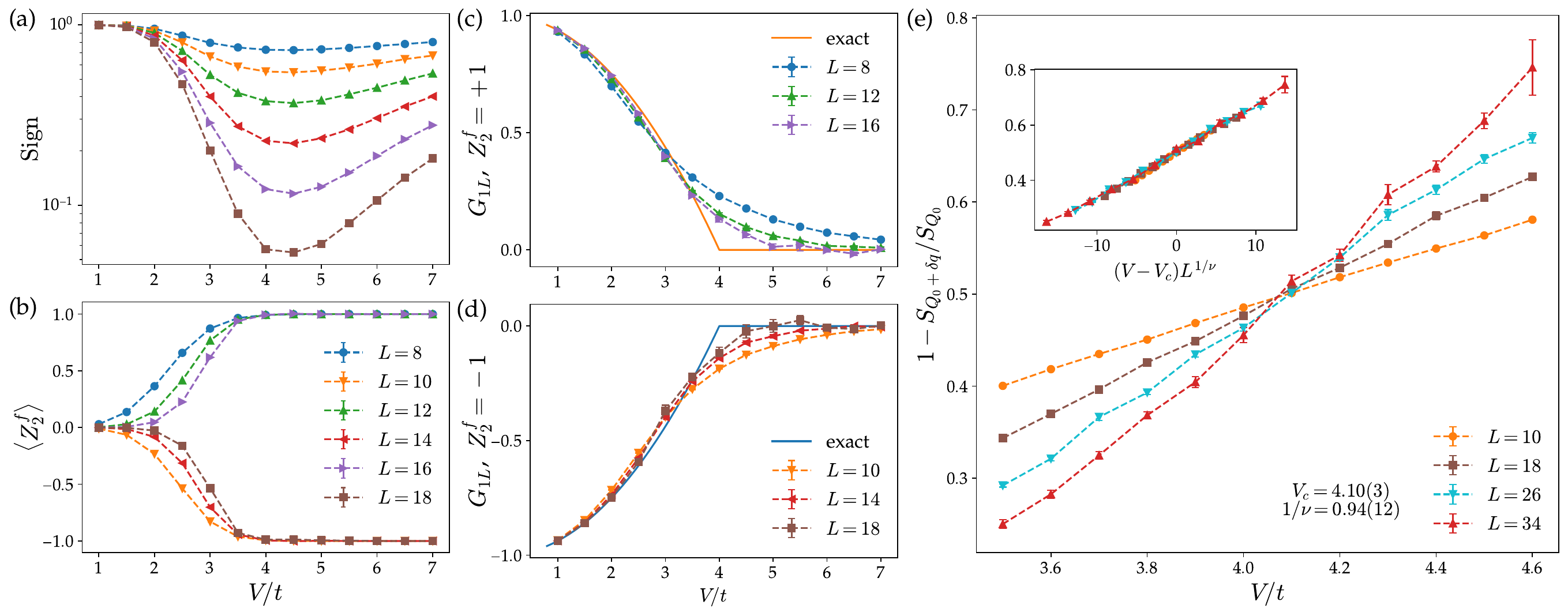}
    \caption{Numerical results on the interacting Kitaev chain at $\Delta = t$. Recognizing the dynamical critical exponent of $z=1$ for the quantum phase transition in this model, we maintain $\beta \propto L$ in all the simulations. Specifically, for subfigures (a), (b), (c) and (d), the inverse temperature is set as $\beta t = L$, while for (e) $\beta t = 0.1L$ to access larger system sizes. The Trotter time step is fixed as $\Delta \tau = 0.1$. (a) displays the average sign. The sign problem is most severe near the quantum critical point $V_c = 4$. (b) The expectation value of the fermion parity $Z^f_2$ operator. In the CDW phase, $Z^f_2$ converges to $(-1)^{L/2}$ since the system is at half-filling. While in the TSC phase, $Z^f_2$ approaches $0$ in the thermodynamic limit, reflecting the degenerate ground state with $Z^f_2 = \pm 1$.  (c)(d) compares the numerical results for the edge correlation function $G_{1L}$ and the analytical solution in the limit $L \to \infty$. In both sectors with $Z^f_2=\pm 1$, there is a good correspondence between numerical and theoretical results. (e) Finite size scaling of the ratio $R = 1 - S_{Q_0 + \delta q} / S_{Q_0}$, where $S_{Q}$ is the CDW structure factor defined as $\frac{1}{L^2} \sum_{i, j} e^{\ima Q (i-j)} \braket{(n_i - 1/2) (n_j - 1/2)}$, $Q = \pi$ and $\delta q = 2\pi / L$.  Inset: Data collapse of $R$ using the critical value $V_c$ and the exponent $ \nu $ extracted from the data of $L = 18, 26, 34$. }
    \label{fig:chain-sign-z2}
\end{figure*}

It is important to note that when $\det\left( I + B \right) = 0$, the matrix inverse in Eq.~\ref{eq:productRuleMain} is singular, and this representation
breaks down. In other words, nearly all elements in the spin group can be effectively handled, except for those with vanishing trace $\trace{\mathcal{O}} = 0$. 
Since the trace is nothing but the Boltzmann weight, a practical Monte Carlo simulation shall automatically avoid any configuration with zero trace,
akin to navigating between ice sheets without encountering any cracks, as depicted in Fig.~\ref{fig:PfQMC1}.

A closed-form formula for the weight $w(\{\phi_n\})$ can be given by:
\eq{
\begin{aligned}
    & \trace{ e^{-\frac{1}{4} \bm \gamma^T h_1 \bm \gamma} \cdots e^{-\frac{1}{4} \bm \gamma^T h_{L_\tau} \bm \gamma} } = \left[(-1)^{NL_\tau/2} 2^N \prod_{i=1}^{L_\tau} \eta(h_i) \right]  \\ 
        & \times\pfaffian \begin{pmatrix}
            \tanh \left( \frac{h_1}{2} \right) & -I & I & \cdots \\
            I & \tanh \left( \frac{h_2}{2} \right) &  \ddots & \vdots \\
             -I  & \ddots & \ddots &  -I \\
            \vdots & \cdots & I & \tanh \left( \frac{h_{L_\tau}}{2} \right)
    \end{pmatrix}.
         \\
\end{aligned}
}{traceToPfaffian}
Here $L_\tau$ needs to be even. This equation bears resemblance to the ``space-time formulation'' of the DQMC algorithm \cite{santos_introduction_2003}, and can be interpreted as an exact discretized version of the Majorana path integral \cite{shankar2017quantum}. This formulation complements the previous iterative formula and is particularly valuable in discussions concerning the sign-problem-free property
(see Supplementary Material for further discussions).

\textit{Pfaffian quantum Monte Calro (PfQMC).}---With the sign ambiguity resolved, we are now equipped to generalize the DQMC algorithm.
Clearly, using either the iterative approach or the closed-form formula, we can calculate the Boltzmann weight and apply Markov chain Monte Carlo methods, such as the Metropolis-Hastings algorithm. However, the efficiency in such naive implementation will be extremely poor, as each update would require recalculating the weight from scratch.
This issue is addressed by the fast update algorithm in DQMC, and similar technique is also discovered for PfQMC. 

Suppose after one local update, the FGO on time slice $n$ is changed according to $e^{-\frac{1}{4} \bm \gamma^T h_n \bm \gamma} \to e^{-\frac{1}{4} \bm \gamma^T m \bm \gamma} e^{-\frac{1}{4} \bm \gamma^T h_n \bm \gamma} $, where $m$ is some low-rank matrix. 
Using Eq.~(\ref{eq:productRuleMain}), the acceptance ratio is given by 
\eq{
        R = \frac{ \trace{ e^{ -\frac{1}{4} \bm \gamma^T m \bm \gamma} \mathcal O_n }}{ \trace{\mathcal O_n} }
        = (-1)^N \eta_m \pfaffian \begin{pmatrix}
            G_m & -I \\ I & G \end{pmatrix},
}{AcceptanceRatio}
where $\mathcal O_n$ represents the cyclic product of FGOs, namely $\mathcal O_n = e^{ -\frac{1}{4} \bm \gamma^T h_n \bm \gamma} \cdots e^{ -\frac{1}{4} \bm \gamma^T h_{L_\tau} \bm \gamma} e^{ -\frac{1}{4} \bm \gamma^T h_1 \bm \gamma} \cdots e^{ -\frac{1}{4} \bm \gamma^T h_{n-1} \bm \gamma}$. $G$ is the Green's function associated with $\mathcal O_n$. $G_m = \tanh(m/2)$ shares the same rank with $m$. Using the Pfaffian identity
$\pfaffian{\left[A + B C B^T\right]} = \pfaffian{\left[A\right]} \times \frac{\pfaffian\left[ C^{-1} + B^T A^{-1} B \right]}{\pfaffian\left[C^{-1}\right]}$,
the evaluation of the ratio requires only the Pfaffian of some low-rank matrices, which can be computed with $O(1)$ time complexity. A practical implementation of PfQMC \footnote{A code implementation can be found at \url{https://github.com/zyHan2077/PfQMC}} mirrors DQMC in every aspects, especially the computational complexity. However, the iterative evaluation of the sign as mentioned earlier results in an additional constant overhead. This overhead can be suppressed by leveraging the fast update algorithm's ability to track the sign during the updating process, while the iterative method is used only to ensure the numerical stability (see Supplemental Materials for examples).

\textit{Applications to the interacting Kitaev chain.}---We consider the Kitaev chain \cite{Kitaev_2001} with nearest-neighbor density-density interaction $V$. This model can be exactly solved when $\Delta = t$ \cite{miao2017exact}, but due to the absence of a conserved $\text{U}(1)$  charge, it has not been treated by any DQMC-based methods before. For a chain with length $L$, the Hamiltonian is given by:
\begin{equation}
    H= \sum_{j=1}^{L-1} [ (-t c^\dagger_{j} c_{j+1} + \Delta c^{\dagger}_{j+1} c^\dagger_{j} + \text{h.c.} ) +  V n_{j}n_{j+1} ].
\end{equation}
Here an open boundary condition is assumed. When $0 < V < V_c = 4t$, the ground state is in the topological superconductor (TSC) phase with two-fold degeneracy, while for $V > V_c$, the ground state is a charge density wave (CDW) state \cite{miao2017exact}.

In our PfQMC simulation, HS transformation in the Majorana hopping channel \cite{Li2015PRB} is adopted. 
The sign problem is most severe around the quantum critical point $V_c$, consistent with previous research \cite{mondaini2022quantum}.
Although the sign exhibits a typical exponential decaying feature, hindering us from exploring larger systems and lower temperatures, we can still utilize results from small system sizes to study the phase transition and benchmark our algorithm.

In TSC phase, the ground state degeneracy can be distinguished by the fermion number parity operator $Z^f_2 = (-1)^{\hat{N}}$, with $\hat{N} = \sum_j c^\dagger_j c_j$ the total fermion number. 
The PfQMC result for the finite-temperature expectation value $\braket{Z^f_2}$, evaluated using standard Wick's theorem decomposition \cite{balian_nonunitary_1969}, is depicted in Fig.~\ref{fig:chain-sign-z2}(b). In the thermodynamic limit, $\braket{Z^f_2} \to 0$ in the TSC phase, signaling the ground state degeneracy. In the CDW phase, $\braket{Z^f_2} \to (-1)^{L/2}$.

The exact solution \cite{miao2017exact} also predicts that the \textit{edge correlation function}, defined by $G_{1L} = \braket{\ima \gamma^1_1 \gamma^2_L}$, shall behave as $\lim_{L \to \infty} G_{1L} = (-1)^{Z^f_2} \left[ 1 - (V/4t)^2 \right]$ in the TSC phase, and shall vanish in the CDW phase. By performing a projection onto the two symmetry sectors with $Z^f_2 = \pm 1$, we can get numerical results in accordance with this exact solution, as depicted in Fig.~\ref{fig:chain-sign-z2}(c)(d). The CDW transition is also studied. Fig.~\ref{fig:chain-sign-z2}(e) shows the finite size scaling \cite{melchert2009autoscalepyprogramautomatic} result, revealing a critical interaction $V_c \approx 4.1$, which is close to the exact result of $V_c = 4$. A critical exponent $\nu \approx 1.0$ is also obtained, in accordance with the 2d Ising universality class.

Away from half-filling, where an exact solution is absent, PfQMC can still be employed. 
The quantum phase transition between the CDW phase and the incommensurate CDW (ICDW) phase \cite{miao2018majorana} is studied, as detailed in the Supplemental Materials. These results convincingly demonstrate the ability of PfQMC to perform unbiased simulations of \textit{general} fermionic Hamiltonian.

\textit{Majorana-resolved sign.}---As another application of PfQMC, we consider
an intriguing possibility presented by this new formalism: the Majorana-resolved sign.

\begin{figure}[t]
    \includegraphics*[width=\linewidth]{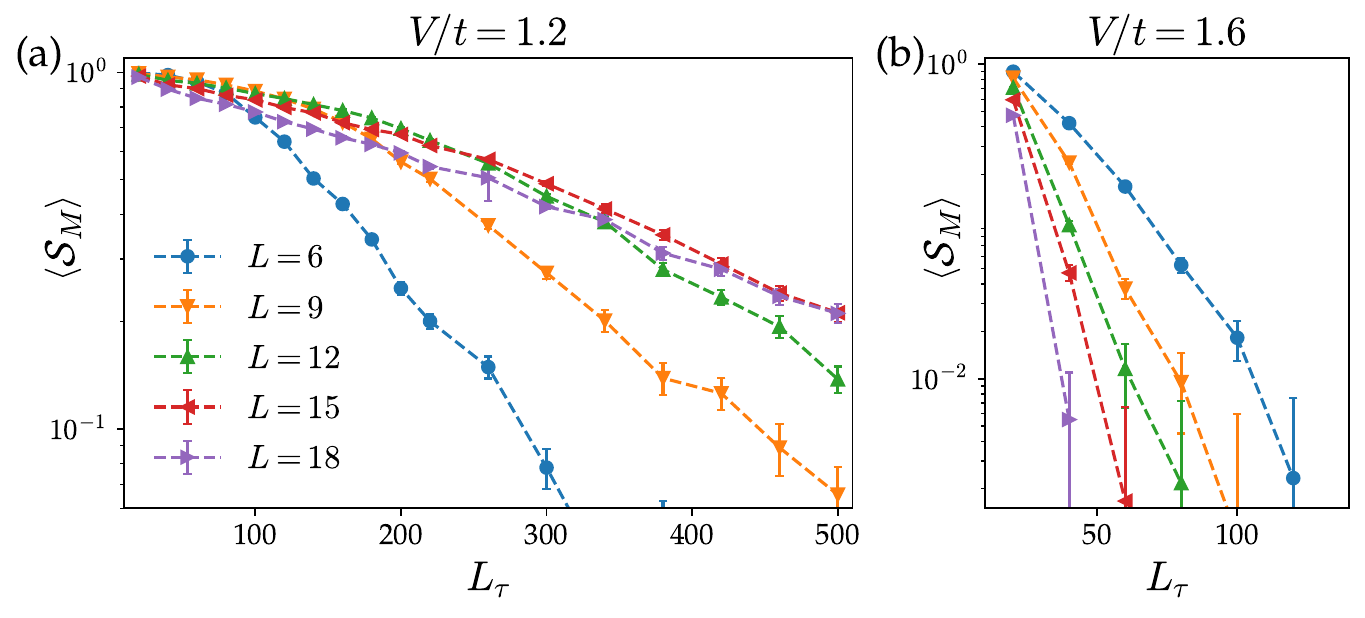}
    \caption{Dependence of the Majorana-resolved sign $\braket{\mathcal{S}_{M}}$ on the system size $L$ and imaginary time steps $L_\tau$. 
    The calculation is for the spinless $t$-$V$ model at half-filling and the Trotter time step is fixed as $\Delta\tau = 0.1$.
    (a) For $V=1.2t < V_c$, the sign reduces with increasing system size at high temperatures, while at low enough temperatures this behavior reverses. 
    (b) For $V=1.6t>V_c$, the sign exhibits a rapid decay with increasing $L_\tau$ and system size $L$.  }
    \label{fig:resolved-sign-main}
\end{figure}

The Majorana-resolved sign is a generalization of the recently introduced “spin-resolved sign” \cite{mondaini2022quantum, mondaini2023universality}, a numerical object reminiscent of the sign in the sign problem. This ``resolved sign'' can in principle be defined whenever the model falls into the Kramers or Majorana sign-problem-free classes \cite{Li2016PRL}. However, previous studies based on DQMC can only address this issue for models belonging to the Kramers class. Thanks to PfQMC, studying the Majorana class has become possible. Here we examine the spinless $t$-$V$ model on a honeycomb lattice \cite{li2015fermion, Huffman2014PRB, wang2014fermionic, wang2016stochastic}, with its Hamiltonian given by $H = -t \sum_{\braket{ij}} (c^\dagger_i c_j + \text{h.c.}) + V \sum_{\braket{ij}} n_i n_j$. Using a symmetric HS decoupling scheme with respect to the two Majorana flavors \cite{Li2015PRB}, the partition function becomes $Z = \sum_c w^2(c)$, and the Majorana-resolved sign is defined as $\braket{\mathcal{S}_{M}} = \sum_c \mathrm{sgn}[w(c)] w^2(c) / \sum_c w^2(c)$, where $c$ denotes the auxiliary-field configuration. As displayed in Fig.~\ref{fig:resolved-sign-main}, this resolved sign displays a usual exponentially decaying feature in the CDW phase at $V \gtrsim V_c \approx 1.36$, while a non-trivial behavior is found in the semi-metal phase with $V \lesssim V_c$, where it grows with increasing lattice size at low enough temperatures. See also the Supplemental Materials for details. These results suggest this Majorana-resolved sign does contain information about the quantum phase transition of this model.

{\it Discussions and concluding remarks.}---The framework of PfQMC proposed in this work serves as a natural and nontrivial extension of DQMC, and brings DQMC towards its full potential. This enables QMC to study the interacting Kitaev chain for the first time, and to observe non-trivial behavior of the Majorana-resolved sign.

Although in the present work we have only demonstrated the correctness and applicability of PfQMC for finite temperature calculations, extending its application to any DQMC-related methods, including the projector QMC or the constrained-path QMC, should encounter no obvious obstacles. We believe that PfQMC could provide promising route to address challenging issues such as the nature of the ground state of the Hubbard model 
\cite{Arovas_2022, Qin_2022} and the mechanism of high-temperature superconductivity \cite{Dagotto1994RMP, wen2006doped, Scalapino2012RMP, keimer_quantum_2015, Fradkin-Kivelson-Tranquada2015RMP, Varma2020}.

We would also like to note that the Grassmann representation utilized here is also 
importantly developed in the context of matchgate circuits \cite{valiant_quantum_2001}, often referred to as the fermionic linear optics \cite{terhal_classical_2002, knill_fermionic_2001, bravyi_lagrangian_2004, Wan_2023, Bravyi_2017} within the quantum computing community. The interplay between PfQMC and matchgates is yet to be further explored, and we believe that important questions may be addressed along this line in the future, such as the classical simulability of quantum circuits and the representation power of fermionic Gaussian states. 

{\it Acknowledgement}: We sincerely thank Jinmin Wang, Yiyang Wang, Zheng-Zhi Wu, Ning Xia, and Shi-Xin Zhang for helpful discussions, and especially Zi-Xiang Li for related collaborations. This work is supported in part by NSFC under Grant Nos. 12347107 and 12334003 (ZYH, ZQW, and HY), by MOSTC under Grant No. 2021YFA1400100 (HY), and
by the Xplorer Prize through the New Cornerstone Science Foundation (HY).

%

\appendix

\begin{widetext}
	\appendix
	\section*{Supplemental Materials}
	\setcounter{equation}{0}
	\setcounter{figure}{0}
	\setcounter{table}{0}
	\makeatletter
	\renewcommand{\theequation}{S\arabic{equation}}
	\renewcommand{\thefigure}{S\arabic{figure}}
	\renewcommand{\bibnumfmt}[1]{[S#1]}
	\renewcommand{\citenumfont}[1]{S#1}
	\renewcommand{\thesubsection}{\Alph{subsection}}
	\subsection{A. Grassmann algebra, Clifford algebra, Pfaffian, and fermionic coherent states}
	
	For sake of completeness, we briefly review the Grassmann algebra and its Gaussian integrals. We also introduce fermion coherent states and its completeness relation.
	
	Grassmann algebra $\mathcal{G}_n$ is generated by formal variables $\vect \theta =\{\theta_1, \cdots, \theta_n\}$ with properties of $\{\theta_i,\theta_j\}=0$. 
	The elements in $\mathcal{G}_n$, which are called Grassmann numbers,  can be represented as complex coefficients polynomials of the Grassmann variables $\alpha + \sum_{p=1}^{n} \sum_{1\leq a_1< \cdots <a_p\leq n} \alpha_{a_1\cdots a_p} \theta_{a_1}\cdots\theta_{a_p}$. We may call this a canonical form of the Grassmann number. The Clifford algebra $\mathcal{C}_n$, generated by Majorana operators $\bm \gamma = \{\gamma_1, \cdots \gamma_n \}$, can be viewed as a ``quantization'' of the Grassmann algebra, for their only difference lies in the commutation relation $\{\gamma_i, \gamma_j\} = 2\delta_{ij}$. Elements in $\mathcal{C}_n$ can be written as $\alpha + \sum_{p=1}^{n} \sum_{1\leq a_1< \cdots <a_p\leq n} \alpha_{a_1\cdots a_p} \gamma_{a_1}\cdots\gamma_{a_p}$, this can be called the canonical form of Clifford algebra, respectivly. The Grassmann representation \cite{bravyi_lagrangian_2004} is introduced as a natural map $\omega_{\bm \theta}: \mathcal{C}_{n} \to \mathcal{G}_{n}$ defined as follows:
	\begin{equation}
		\omega_{\bm \theta}\left( \alpha + \sum_{p=1}^{n} \sum_{1\leq a_1< \cdots <a_p\leq n} \alpha_{a_1\cdots a_p} \gamma_{a_1}\cdots\gamma_{a_p} \right) = \alpha + \sum_{p=1}^{n} \sum_{1\leq a_1< \cdots <a_p\leq n} \alpha_{a_1\cdots a_p} \theta_{a_1}\cdots\theta_{a_p},
	\end{equation}
	which is a one-to-one linear map. 
	
	Grassmann representation of $X \in \mathcal{C}_{2N}$ is said to be Gaussian (or of Gaussian form), if $\omega_{\bm \theta}(X) = \eta \times \exp \left[ - \frac{1}{2} \bm \theta^T G \bm \theta \right]$, with $\eta \in \mathbb{C}$ and $G$ a $2N \times 2N$ skew-symmetric matrix. Hereafter, $\bm \theta$ is always assumed to be a column vector $\bm \theta = (\theta_1, \theta_2, \cdots \theta_{2N})^T$ for simplicity.

	The partial derivatives on the Grassmann algebra is defined as 
	$\frac{\partial}{\partial \theta_i}1 =0,\  \frac{\partial}{\partial \theta_i}\theta_j =\delta_{ij},\  \left\{\frac{ \partial}{\partial \theta_i},\frac{ \partial}{\partial \theta_j} \right\} = 0$, 
	together with the Leibniz's rule $\frac{\partial}{\partial \theta_i} (\theta_j f(\theta))=\delta_{ij}f(\theta)-\theta_j\frac{\partial}{\partial \theta_j}f(\theta)$. 
	The integration in this algebra is defined to be equivalent to the partial derivative $\int d\theta_i\equiv\frac{\partial}{\partial \theta_i}$, and we adopt the notation that 
	$\int D\vect\theta \equiv \int d\theta_n\cdots \int d\theta_2\int d\theta_1$.
	Note that the variable can be transformed by $\vect{\theta}' = T \vect{\theta},\  T \in \text{GL} (n,\mathbb{C})$, the integral measure is changed accordingly that $\D\vect \theta'=\det (T)^{-1}\D \vect \theta $.
	With these definitions and properties, an elegant formula can be deduced for the Gaussian integrals in Grassmann algebra:
	\eq{
	\int D\vect \theta \exp\left(\frac 1 2 \vect\theta^T M \vect\theta\right) = \pfaffian \left(M\right),
	}{Grassmann_Gaussian_Integral}
	where $\pfaffian\left( M \right)$ is the Pfaffian of a complex anti-symmetric matrix $M$. Another Gaussian integral that will be extensively used is:
	\begin{equation}
		\int D\vect \theta \exp\left(\bm\xi^T \bm \theta +  \frac 1 2 \vect\theta^T M \vect\theta\right) = \pfaffian \left(M\right) \times \exp\left(\frac{1}{2} \bm \xi^T M^{-1} \bm\xi \right).
	\end{equation}
	
	By extending the Hilbert space from complex numbers to Grassmann numbers, we can introduce the fermion coherent state $\ket{\vect \xi} = \exp (-\sum_i\xi_i \hat c^\dagger_i)\ket{0}$, which is an eigenstate of the annihilation operators $c_i$, since $c_i\ket{\vect \xi} = \xi_i\ket{\vect \xi}$, here $\bm \xi$ is an array of Grassmann variables. We may also introduce conjugate Grassmann variables and adjoint coherent state $\bar{\bm \xi}$ and $\bra{\bar{\bm\xi}}$, defined by $\bra{\bar{\bm\xi}} = \bra{0} \exp(\sum_i \bar \xi_i c_i)$, with $\bra{\bar{\bm \xi}} c_i^\dagger =\bra{\bar{\bm\xi}} \bar{\xi_i}$ \cite{altland2010condensed}. 
	
	The fermion coherent states are over-complete basis of the Hilbert space and the completeness relation is given by
	\eq{
	\int D (\bar{\bm \xi}, \bm\xi) \exp \left (-\sum_j \bar{\xi}_j \xi_j\right) \ket{\vect \xi} \bra{\bar{\vect \xi}} = 1.
	}{completeness}
	here $D (\bar{\bm \xi}, \bm\xi) = \prod_i \text{d} \bar{\xi}_i \xi_i$. In the following derivations, we often have to switch between $D (\bar{\bm \xi}, \bm\xi)$ and the integration measure $D \bm \Xi$, where $\bm \Xi = (\xi_1, \cdots \xi_n, \bar{\xi}_1, \cdots \bar{\xi}_n)^T$. Just keep in mind that a $(-1)^{\frac{n(n-1)}{2}}$ prefactor needs be included whenever one performs such switch. Specifically, when $n=2N$, this prefactor becomes $(-1)^{N(2N-1)} = (-1)^N$.

	\subsection{B. Proof of the main results}
	
	First, we prove that if $X_1, X_2 \in \mathcal{C}_{2N}$ each have a Gaussian Grassmann representation, then the Grassmann representation of their product is still of Gaussian form. 
	
	Let $\omega_{\bm \theta}(X_i) = \eta_i \times e^{-\frac{1}{2} \bm \theta^T G_i \bm \theta}$, with $i = 1, 2$. We enlarge the Hilbert space by considering $2N$ fermionic degree of freedoms (d.o.fs), labeled by $c_i, c^\dagger_i, i \in \{1, 2, \cdots 2N\}$. Note that these $c$'s will only serve as a mathematical construction, and are not related to the original physical fermionic d.o.fs. We use $\omega(X, c^\dagger + c)$ to denote a change of variable where each $\gamma_i$ in the canonical form of $X$ is substituted by $c^\dagger_i + c_i$, i.e.,
	\begin{equation}
		\omega\left( \alpha + \sum_{p=1}^{2N} \sum_{1\leq a_1< \cdots <a_p\leq 2N} \alpha_{a_1\cdots a_p} \gamma_{a_1}\cdots\gamma_{a_p}, c^\dagger + c \right) = \alpha + \sum_{p=1}^{2N} \sum_{1\leq a_1< \cdots <a_p\leq 2N} \alpha_{a_1\cdots a_p} (c^\dagger_{a_1} + c_{a_1}) \cdots (c^\dagger_{a_p} + c_{a_p}),
	\end{equation}
	
	We now have
	\eq{
	\begin{aligned}
			&\bra{\bar{\bm\psi}} \omega(X_1, c^\dagger + c) \omega(X_2, c^\dagger + c) \ket{\bm\psi} \\
			= & \int D (\bar{\bm\theta}, \bm \theta) \, e^{-\bar{\bm\theta}^T \bm{\theta}}  \bra{\bar{\bm\psi}} \omega(X_1, c^\dagger + c) \ket{\bm\theta} \bra{\bar{\bm\theta}} \omega(X_2, c^\dagger + c) \ket{\bm\psi} \\
			= & \int D (\bar{\bm\theta}, \bm \theta) \, e^{-\bar{\bm\theta}^T \bm{\theta}}   \omega(X_1, \bar{\bm\psi} + \bm\theta) \omega(X_2, \bar{\bm\theta} + \bm\psi) e^{\bar{\bm\psi}^T \bm \theta} e^{\bar{\bm\theta}^T \bm \psi}  \\
			= & \eta_1 \eta_2 \exp\left[
				-\frac{1}{2} \begin{pmatrix}
					\bar{\bm\psi}^T & \bm\psi^T
				\end{pmatrix} \begin{pmatrix}
					G_1 & 0 \\
					0 & G_2
				\end{pmatrix} \begin{pmatrix}
					\bar{\bm\psi} \\ \bm\psi
				\end{pmatrix}
			\right] \times \\
			& \hspace*{2cm} \int D (\bar{\bm\theta}, \bm \theta) \exp\left[-\frac{1}{2} \begin{pmatrix}
				\bm\theta^T & \bar{\bm\theta}^T
			\end{pmatrix} \begin{pmatrix}
				G_1 & -I \\
				I & G_2
			\end{pmatrix} \begin{pmatrix}
				\bm\theta \\ \bar{\bm\theta}
			\end{pmatrix} - \begin{pmatrix}
				\bar{\bm\psi}^T (G_1-I) & \bm\psi^T (G_2+I) 
			\end{pmatrix} \begin{pmatrix}
				\bm\theta \\ \bar{\bm\theta}
			\end{pmatrix} \right]  \\
			= & (-1)^N \eta_1 \eta_2 \pfaffian \begin{pmatrix}
				G_1 & -I \\
				I & G_2
			\end{pmatrix} \times \exp\left[
				-\frac{1}{2} \begin{pmatrix}
					\bar{\bm\psi}^T & \bm\psi^T
				\end{pmatrix} \begin{pmatrix}
					G_1 & 0 \\
					0 & G_2
				\end{pmatrix} \begin{pmatrix}
					\bar{\bm\psi} \\ \bm\psi
				\end{pmatrix}
			\right]   \\
			&\hspace*{2.8cm} \times  \exp\left[
				\frac{1}{2} \begin{pmatrix}
					\bar{\bm\psi}^T (G_1-I) &  {\bm\psi}^T (G_2 +I) 
				\end{pmatrix} \begin{pmatrix}
					G_2 (I + G_1 G_2)^{-1} & (I+G_2G_1)^{-1} \\
					- (I + G_1 G_2)^{-1} & G_1 (I+G_2G_1)^{-1}
				\end{pmatrix} \begin{pmatrix}
					(G_1+I) \bar{\bm\psi} \\ (G_2-I) {\bm\psi}
				\end{pmatrix}
			\right] \\
			=& (-1)^N \eta_1 \eta_2 \pfaffian \begin{pmatrix}
				G_1 & -I \\
				I & G_2
			\end{pmatrix} \times \\
			&\hspace*{-0.2cm} \exp\left[
				-\frac{1}{2} \begin{pmatrix}
					\bar{\bm\psi}^T & {\bm\psi}^T
				\end{pmatrix} \begin{pmatrix}
					G_1 - (G_1-I) G_2 (I + G_1 G_2)^{-1} (G_1+I)& -(G_1 - I) (I + G_2 G_1)^{-1} (G_2-I) \\
					(G_2 + I)(I + G_1 G_2)^{-1} ({G_1+I}) & G_2 - (G_2 + I) G_1 (I + G_2 G_1)^{-1} (G_2-I)
				\end{pmatrix} \begin{pmatrix}
					\bar{\bm\psi} \\ {\bm\psi}
				\end{pmatrix}
			\right].
			\end{aligned}
	}{eq:grassmanProd}
	By introducing $B_i$ such that $G_i = (I-B_i)/(I+B_i)$, the above expression can be reformulated as %
	\eq{
	\begin{aligned}
			&\bra{\bar{\bm\psi}} \omega(X_1, c^\dagger + c) \omega(X_2, c^\dagger + c) \ket{\bm\psi} \\
			=& (-1)^N \eta_1 \eta_2 \pfaffian \begin{pmatrix}
				G_1 & -I \\
				I & G_2
			\end{pmatrix} \times \exp{\left[-\frac 1 2 (\bar{\bm\psi}^T+\bm\psi^T) \frac{I-B_1B_2}{I+B_1B_2}(\bar{\bm\psi}+\bm\psi)\right]}\times \exp (\bar{\bm\psi}^T\bm\psi).
			\end{aligned}
	}{}
	Comparing with $\bra{\bar\psi} \omega(X_1, c^\dagger + c) \omega(X_2, c^\dagger + c) \ket{\psi} = \bra{\bar\psi} \omega(X_1X_2, c^\dagger + c) \ket{\psi} = \omega(X_1X_2, \bar\psi+\psi) e^{\bar\psi\psi}$, we thus proved the Grassmann representation of $X_3 = X_1 X_2$ is indeed Gaussian, with
	\begin{equation}
		\eta_3 = (-1)^N \eta_1 \eta_2 \pfaffian \begin{pmatrix}
			G_1 & -I \\
			I & G_2
		\end{pmatrix}, \ B_3 = B_1 B_2, \ G_3 = \frac{I - B_3}{I + B_3},
	\end{equation}
	and this gives Eq.~\eqref{eq:productRuleMain} in the main article. It's crucial to note that $B$ can be derived from $G$ through the relation $B = (I-G)/(I+G)$. Accordingly, a Gaussian Grassmann representation can be denoted as $(\eta, B)$, as in the main article.
	
	The Grassmann representation of one certain FGO, say $e^{-\frac{1}{4} \bm \gamma^T h \bm \gamma}$, could then be evaluated. Let $\omega_{\bm\theta}(e^{-\frac{1}{4} \bm \gamma^T x h \bm \gamma}) = \eta(x) \times e^{-\frac{1}{2} \bm \theta^T G(x) \bm \theta}$, We aim to determine $\eta, G$ by solving a differential equation with respect to $x$ over the interval  $[0, 1]$. 
	Using the fact $\omega_{\bm\theta}(e^{-\frac{1}{4} \bm \gamma^T \Delta x h \bm \gamma}) = \omega_{\bm\theta}(1 - \frac{1}{4} \bm \gamma^T \Delta x h \bm \gamma) + O(\Delta x^2) = e^{-\frac{1}{2} \bm \theta^T (\Delta x  h / 2) \bm \theta} + O(\Delta x^2)$, $G(x)$ can be solved easily since $B(x) = \frac{I - G(x)}{I + G(x)}$ obeys the equation $B(x + \Delta x) = B(x) e^{-\Delta x h} $, therefore $B(x)=e^{-xh}, G(x) = \tanh(\frac{xh}{2})$.
	The $\eta(x)$ then satisfies $\eta(x+\Delta x) = (-1)^N \pfaffian \begin{pmatrix}
		\Delta x h/2 & -I \\
		+I & \tanh(xh/2)
	\end{pmatrix} \eta(x) + O(\Delta x^2)$, which gives a differential equation:
	
	\begin{equation}
	\frac{\text{d} \eta}{\eta \text{d} x} = \frac{1}{4} \trace{h \tanh\left(\frac{x h}{2}\right)}, \quad \eta(0) = 1.
	\label{eq:DifferentialEqForEta}
	\end{equation}
	Using this equation, $\eta$ can be solved in closed form as follows:
	\begin{equation}
		\eta(1) = (-1)^N \pfaffian{\begin{pmatrix}
			\sqrt{2} \sinh(\frac{h}{4}) & -I \\
			I & \sqrt{2} \sinh(\frac{h}{4})
		\end{pmatrix}}.
	\end{equation}
	Or, an equivalent formula could be found
	\begin{equation}
		\eta(1) = (-1)^N \det\left(\cosh\left(h/4\right)\right) \pfaffian{\begin{pmatrix}
			\tanh(\frac{h}{4}) & -I \\
			I & \tanh(\frac{h}{4})
		\end{pmatrix}}.
	\end{equation}
	
	Note that the differential equation \eqref{eq:DifferentialEqForEta} actually has singularities whenever $I + B(x) = I + e^{-x h}$ is non-invertible. However, if the restriction of analytical continuity is assumed, the solution is unique and we are thus able to ``cross'' the singularity points.
	
	A trace over product of elements in $\mathcal{C}_{2N}$ can be turned into an integral over their Grassmann representations:
	\eq{
	\trace{X_1 X_2\cdots X_L} = (-1)^{NL/2} \ 2^N  \int D \bm \theta \ \omega_{\bm \theta_1}(X_1) \cdots \omega_{\bm \theta_L}(X_L) \times \exp \left[  \frac{1}{2} \bm \theta^T \begin{pmatrix}
				0 & I & -I & \cdots & I \\
				-I & 0 & I & \ddots & \vdots  \\
				I & -I & 0 & \ddots & \vdots  \\
					\vdots  & \ddots & \ddots & \ddots & I \\
				-I & \cdots & \cdots  & -I & 0
			\end{pmatrix} \bm \theta
				\right], \ \bm \theta^T = \left( \bm \theta_1, \cdots \bm \theta_L \right).
	}{traceToIntegral}
	Note that $L$ must be an even integer here. The $L=2$ case of this formula was initially introduced by S. Bravyi in his seminal work \cite{bravyi_lagrangian_2004}, and its generalization to arbitrary $L$ has recently been known \cite{Bravyi_2017, Wan_2023} to the quantum computation community, particularly in its association with the concept of ``matchgate circuits''. Although there have been previous contributions on this topic, we present an independently derived proof based solely on Gaussian integrals. We believe that our approach will provide greater transparency and insight for the audience.
	
	The trace over fermionic degrees of freedom can be evaluated within the enlarged Hilbert space. Because the Hilbert space dimension increases from $2^N$ to $2^{2N}$, a prefactor of $\frac{1}{2^N}$ must be included.
	Consequently,

	\begin{equation}
		\begin{aligned}
			&\trace{X_1 X_2 \cdots X_L} \\
			&= \frac{1}{2^N} \trace{\omega(X_1, c^\dagger+c) \omega(X_2, c^\dagger+c) \cdots }\\
			&= \frac{1}{2^N} \int \left[ \prod_{i=1}^L D(\bar{\bm\psi_i}, \bm\psi_i) \right] \times \\
			&\hspace*{1cm} \bra{-\bar{{\bm\psi}}_L} \omega(X_1, c^\dagger+c) \ket{{\bm\psi}_1}  e^{-\bar{\bm\psi}^T_1 {\bm\psi}_1}  \bra{\bar{\bm\psi}_1} \omega(X_2, c^\dagger+c) \ket{{\bm\psi}_2} e^{-\bar{\bm\psi}^T_2 {\bm\psi}_2} \cdots  \bra{\bar{\bm\psi}_{L-1}} \omega(X_L, c^\dagger+c) \ket{{\bm\psi}_L} e^{-\bar{\bm\psi}^T_L {\bm\psi}_L} \\
			&=  \frac{1}{2^N} \int \left[ \prod_{i=1}^L D(\bar{\bm\psi_i}, \bm\psi_i) \right] \, \omega(X_1, -\bar{\bm\psi}_L + {\bm\psi}_1) \cdots \omega(X_L, \bar{\bm\psi}_{L-1} + {\bm\psi}_L) \times \exp \left[ - \bar{\bm\Psi}^T M \bm\Psi \right],
		\end{aligned}
		\label{eq:traceDoubled}
	\end{equation}
	with $M = \begin{pmatrix}
		I & -I & & \\
			&  I & -I & \\
			& & \ddots & \ddots \\
		I  & & & I 
	\end{pmatrix}$ and $\bm\Psi = \begin{pmatrix} \bm\psi_1 \\ \vdots \\ \bm\psi_L\end{pmatrix}$. 
	
	Since the fermionic d.o.fs are doubled,  it is necessary to integrate out the uninvolved d.o.fs. This can be achieved by introducing the transformed grassmann numbers $(\bm\theta, \tilde{\bm\theta})$, defined by
	\begin{equation}
		\begin{pmatrix}
			\bm\theta \\ \tilde {\bm\theta}
		\end{pmatrix} = \begin{pmatrix}
			I & & & & & & &  -I \\
				&I& & & I & & &  \\
				& &I& &   & I & & \\
				& & & \ddots &   & & \ddots & \\
				I & & & & & & &  I \\
				&I& & & -I & & &  \\
				& &I& &   & -I & & \\
				& & & \ddots &   & & \ddots & \\
		\end{pmatrix} \begin{pmatrix}
			\bm\Psi \\ \bar{\bm\Psi}
		\end{pmatrix} \equiv U \begin{pmatrix}
			\bm\Psi \\ \bar{\bm\Psi}
		\end{pmatrix}.
	\end{equation}
	Note that each $I$ here in $U$ is a $2N \times 2N$ identity matrix. Consequently, we have $\det(U) = 2^{2NL}$. By rewriting the integral \eqref{eq:traceDoubled} in terms of $(\theta, \tilde \theta)$, we obtain
	\begin{equation}
		\begin{aligned}
			&\trace{X_1 X_2 X_3\cdots} \\
			&=  \frac{2^{2LN}}{2^N} \int D(\bar{\bm\theta}, \bm\theta) \, \omega(X_1, \theta_1) \cdots \omega(X_L, \theta_L) \times \exp \left[ - \frac{1}{2} \begin{pmatrix}
				\Psi & \bar\Psi
			\end{pmatrix} \begin{pmatrix}
				& - M^T \\
				M & \\
			\end{pmatrix} \begin{pmatrix}
				\Psi \\ \bar\Psi
			\end{pmatrix}
				\right] \\
				&=  2^{(2L-1)N} \int D(\bar{\bm\theta}, \bm\theta) \, \omega(X_1, \theta_1) \cdots \omega(X_L, \theta_L) \times \exp \left[ - \frac{1}{2} \begin{pmatrix}
				\theta & \tilde\theta
			\end{pmatrix} (U^{-1})^T \begin{pmatrix}
				& - M^T \\
				M & \\
			\end{pmatrix} U^{-1} \begin{pmatrix}
				\theta \\ \tilde\theta
			\end{pmatrix}
				\right] \\
				&=  2^{(2L-1)N} \int D(\bar{\bm\theta}, \bm\theta) \, \omega(X_1, \theta_1) \cdots \omega(X_L, \theta_L) \times \exp \left[ - \frac{1}{2} \begin{pmatrix}
				\theta & \tilde\theta
			\end{pmatrix}\begin{pmatrix}
					A & - C^T \\
				C & B \\
			\end{pmatrix} \begin{pmatrix}
				\theta \\ \tilde\theta
			\end{pmatrix}
				\right] \\
				&=  2^{(2L-1)N} \int D(\bar{\bm\theta}, \bm\theta) \, \omega(X_1, \theta_1) \cdots \omega(X_L, \theta_L) \times \exp \left[ - \frac{1}{2} \theta^T (A+C^T B^{-1} C) \theta
				\right] \pfaffian(B),
		\end{aligned}
	\end{equation}
	where $A, B, C$ can all be straightforwardly carried out, with $\pfaffian(B) = (-1)^{NL/2} / 2^{2N(L-1)}$. This leads to equation Eq.\eqref{traceToIntegral}. Eq.\eqref{traceToPfaffian} can thus also be easily obtained.
	
	Through similar calculations, we can also establish the well-known result that $e^{-h}$ provides a faithful representation of $e^{-c^\dagger h c}$ in DQMC. Let $X_1, X_2$ denote such FGOs with particle-number conservation, and let its Grassmann representation $\omega(X, \bm\psi)$ be defined similarly as
	\eq{
		\omega\left(\alpha + \sum_{p=1}^{N} \sum_{\substack{1 \leq a_1 < \cdots < a_p \leq N \\ 1 \leq b_1 < \cdots < b_p \leq N }} \alpha_{\substack{a_1 \cdots a_p \\ b_1 \cdots b_p}} c^\dagger_{a_1} \cdots c^\dagger_{a_p} c_{b_1} \cdots c_{b_p}, \bm\psi\right) = \alpha + \sum_{p=1}^{N} \sum_{\substack{1 \leq a_1 < \cdots < a_p \leq N \\ 1 \leq b_1 < \cdots < b_p \leq N }} \alpha_{\substack{a_1 \cdots a_p \\ b_1 \cdots b_p}} \bar{\psi}_{a_1} \cdots \bar{\psi}_{a_p} \psi_{b_1} \cdots \psi_{b_p},
	}{grassmannDefDQMC}
	therefore $\braket{\bar{\bm\psi}| X | \bm\psi} = \omega(X, \bm\psi) e^{\bar{\bm\psi}^T\bm\psi}$. If $\omega(X_i, \bm\psi) = e^{\bar{\bm\psi}^T (B_i - 1) \bm\psi}$, then the Grassmann representation of $X_1 X_2$ can be evaluated as:
	
	\begin{equation}
		\begin{aligned}
			\omega(X_1 X_2, \bm\psi) &= \braket{\bar{\bm\psi} |  X_1 X_2 | \bm \psi} e^{-\bar{\bm\psi}^T \bm\psi} \\
			&= e^{-\bar{\bm\psi}^T \bm\psi} \int D(\bar{\bm\theta}, \bm\theta) \, e^{-\bar{\bm\theta}^T \bm\theta} \bra{\bar{\bm\psi}}  X_1 \ket{\bm\theta} \bra{\bar{\bm\theta}} X_2 \ket{\bm \psi} \\
			&= e^{\bar{\bm\psi}^T (B_1 B_2 - 1) \bm\psi}.
		\end{aligned}
	\end{equation}
	Also note that for an FGO $X = e^{- \Delta x c^\dagger h c}$ with $|\Delta x| \ll 1$, we have $\omega(X, \bm\psi) = \exp \left[{\bar{\bm\psi}^T (e^{-\Delta x h} - 1) \bm\psi}\right] + O(\Delta x^2)$. It is thus evident that $e^{-h}$'s form a faithful representation of the particle-number conserving FGOs, therefore the FGOs encountered in DQMC can always be represented by non-invertible matrices.
	
	\subsection{C. Observables and fast updating scheme}
	
	Once given an auxiliary field configuration $\left\{ \phi_n \right\}$, an FGO $X_{\left\{ \phi_n \right\}} = \prod_n e^{-\frac{1}{4} \bm \gamma^T h(\phi_n) \bm \gamma}$ is determined. If the Grassmann representation of $X_{\left\{ \phi_n \right\}}$ is obtained as $\omega_{\bm\theta}(X_{\left\{ \phi_n \right\}}) = \eta \cdot \exp\left[-\frac{1}{2} \bm \theta^T G \bm \theta\right]$, the statistical weight is just
	\begin{equation}
		\trace{X_{\left\{ \phi_n \right\}}} = 2^N \eta,
	\end{equation}
	which can be readily seen by setting $L=2$, $X_1 = X_{\left\{ \phi_n \right\}}$ and $X_2 = I$ in Eq.~\eqref{traceToIntegral}.
	
	Owing to Wick's theorem, physical observables factorize into Majorana Green's functions, denoted as $ g_{ij} \equiv \braket{\gamma_i \gamma_j}$ \cite{balian_nonunitary_1969}. The Majorana bilinear $\gamma_i \gamma_j$ adopts a Grassmann representation, $\omega_{\bm \theta}(\gamma_i \gamma_j) = e^{-\frac{1}{2} \bm \theta^T \Gamma(i,j) \bm \theta} - 1$, where the matrix $\Gamma$ has only non-zero off-diagonal elements $[\Gamma(i,j)]_{ji} = -[\Gamma(i,j)]_{ij} = 1$. The Green's function is therefore
	\eq{
	g^\sigma_{ij} = \frac{\trace{\gamma_i \gamma_j  X_{\left\{ \phi_n \right\}}}}{\trace{ X_{\left\{ \phi_n \right\}}}} = \left[G\right]_{ij}.
	}{MajoranaGreenResult}
	This result can be derived by applying Eq.~\eqref{traceToIntegral}. Hence $G$ is just the Green's function associated with the FGO.
	
	The success of DQMC in large-scale simulations has largely relied on the fast update scheme \cite{blankenbecler1981monte,scalapino1981monte, sato2021prb} for local updates. Suppose after the update, the FGO on time slice $n$ is changed according to $e^{-\frac{1}{4} \bm \gamma^T h_n \bm \gamma} \to e^{-\frac{1}{4} \bm \gamma^T m \bm \gamma} e^{-\frac{1}{4} \bm \gamma^T h_n \bm \gamma} $, where $m$ is some low-rank matrix. The acceptance ratio is then given by $|R|$, where
	\eq{
		\begin{split}
			R &= \frac{ \trace{ e^{ -\frac{1}{4} \bm \gamma^T m \bm \gamma} X_n }}{\trace{I \cdot X_n} }  =\frac{(-1)^N 2^N \eta_m \eta(X_n) \pfaffian \begin{pmatrix}
				G_m & -I \\ I & G \end{pmatrix}}{(-1)^N 2^N \eta(X_n) \pfaffian \begin{pmatrix}
				0 & -I \\ I & G \end{pmatrix}}\\ 
			&=  \eta_m \frac{\pfaffian \begin{pmatrix}
				G_m & -I \\ I & G \end{pmatrix}}{\pfaffian \begin{pmatrix}
				0 & -I \\ I & G \end{pmatrix}}.
		\end{split}
	}{AcceptanceRatio}
	Here, $X_n$ represents the cyclic product of FGOs, namely $X_n = e^{ -\frac{1}{4} \bm \gamma^T h_n \bm \gamma} \cdots e^{ -\frac{1}{4} \bm \gamma^T h_{N_\tau} \bm \gamma} e^{ -\frac{1}{4} \bm \gamma^T h_1 \bm \gamma} \cdots e^{ -\frac{1}{4} \bm \gamma^T h_{n-1} \bm \gamma}$. $G$ is the Green's function associated with $X_n$. $G_m = \tanh(m/2)$ shares the same rank with $m$. Using the following Pfaffian identity:
	\eq{
	\frac{\pfaffian{\left[A + B C B^T\right]}}{\pfaffian{\left[A\right]}} =  \frac{\pfaffian\left[ C^{-1} + B^T A^{-1} B \right]}{\pfaffian\left[C^{-1}\right]},
	}{pfaffianLowRank}
	we can set $A = \begin{pmatrix}
		0 & -I \\ I & G
	\end{pmatrix}, \ BCB^T = \begin{pmatrix}
		G_m & 0 \\ 0 & 0
	\end{pmatrix}$ with $C$ an invertible matrix obeying $\operatorname*{dim} C = \operatorname*{rank} G_m = \operatorname*{rank} m $. Also using $\pfaffian\left[A\right] = (-1)^N$ and $A^{-1} = \begin{pmatrix}
		G & I \\ -I & 0
	\end{pmatrix}$, $R$ is converted to the ratio of two Pfaffians for low-rank matrix, which could be calculated in $O(1)$ time-complexity. 
	
	In the following section, we will demontrate the fast update algorithm for the spinless $t$-$V$ model. Notably, the fast update procedure for any model remains identical in PfQMC, and the delay update technique \cite{sun2023delayupdatedeterminantquantum} could also be directly applied.
	
	\subsection{D. Applications to the spinless $t$-$V$ model}

	The spinless $t$-$V$ model is described by the Hamiltonian
	\begin{equation}
		\begin{split}
			H &= H_0 + H_{int} \\
			&= -\sum_{ij} t_{ij} c^\dagger_i c_j + \text{h.c.} + \sum V_{ij} (n_i - 1/2)(n_j - 1/2).
		\end{split}
	\end{equation}
	
	Using Majorana operators, this model can be reformulated as
	\begin{equation}
		\begin{split}
			& H_0 = \sum_{\braket{ij}} \frac{\ima t}{2} \left( \gamma^1_i \gamma^1_j + \gamma^2_i \gamma^2_j \right) \\
			& H_{int} = -\frac{V}{4} \sum_{\braket{ij}} (\ima \gamma^1_i \gamma^1_j) (\ima \gamma^2_i \gamma^2_j).
		\end{split}
	\end{equation}
	For repulsive $V$, the HS transformation can be formulated in a symmetric manner with respect to two Majorana flavours,
	\begin{equation}
		e^{\frac{\Delta \tau V}{4} (\ima \gamma^1_i \gamma^1_j) (\ima \gamma^2_i \gamma^2_j)} = \frac{1}{2} \sum_{\sigma_{ij} = \pm 1} e^{\frac{\lambda}{2} \sigma_{ij} (\ima \gamma^1_i \gamma^1_j + \ima \gamma^2_i \gamma^2_j) - \frac{\Delta\tau V}{4}}.
		\label{eq:HSforSymMajorana}
	\end{equation}
	
	We perform the MCMC update by sweeping through the space-time lattice. When arriving at imaginary time slices $l$ with auxiliary field $\sigma_{ij}$, the weight is 
	\begin{equation}
		w_l = \text{Tr}\left[ e^{-\frac{1}{4} \gamma^T h_{l-1} \gamma} \cdots e^{-\frac{1}{4} \gamma^T h_{1} \gamma} e^{-\frac{1}{4} \gamma^T h_{L} \gamma} \cdots e^{-\frac{1}{4} \gamma^T h_{l} \gamma} \right].
	\end{equation}
	And flipping an auxiliary field $\sigma_{jk}$ results in a change $h_l \to h_l + m$.
	
	Let one first consider the case where $m$ is a rank-2 skew-symmetric matrix. Note that all other cases can be reduced to this minimal scenario. Here $m_{j, k} = 2\ima \lambda \sigma = - m_{k, j}$, the acceptance ratio is
	\begin{equation}
		\begin{split}
			R &= \frac{w'_l}{w_l} = \frac{
			\text{Tr}\left[ e^{-\frac{1}{4} \gamma^T h_{l-1} \gamma} \cdots  e^{-\frac{1}{4} \gamma^T h_{l} \gamma} e^{-\frac{1}{4} \gamma^T m \gamma} \right]
		}{\text{Tr}\left[ e^{-\frac{1}{4} \gamma^T h_{l-1} \gamma} \cdots  e^{-\frac{1}{4} \gamma^T h_{l} \gamma} \right]} \\
		& = (-1)^N \eta_m \pfaffian \begin{pmatrix}
			G_m & -I \\ I & G_l \end{pmatrix} \\
		& = \eta_m \frac{\pfaffian\left[ C^{-1} + B^T X^{-1} B \right]}{\pfaffian \left[ C^{-1} \right]} \\
		&= \eta_m \left[1 - \ima \tanh\left(\lambda\sigma\right) G_{j, k} \right].
		\end{split}
	\end{equation}
	Here $\eta_m = \cosh(\lambda)$ and $C = \begin{pmatrix}
		0  & \ima \sigma \tanh(\lambda) \\
		-\ima \sigma \tanh(\lambda) & 0
	\end{pmatrix}$ represents the rank-2 matrix of $\tanh(m/2)$. $B$ is a $4N \times 2$ matrix which should be chosen such that $B C B^T = \begin{pmatrix}
		G_m & 0 \\ 
		0 & 0
	\end{pmatrix}$. One simple choice is setting $B_{j, 1} = B_{k, 2} = 1$, with all other matrix elements zero.

	For the spinless $t$-$V$ model here, each auxiliary field couples to $4$ Majorana fermions, therefore $m$ is a rank-4 skew-symmetric matrix, with $m_{(1j), (1k)} = +2 \ima \lambda \sigma, \ m_{(2j), (2k)} = + 2 \ima \lambda \sigma$. The acceptance ratio $R$ can be evaluated by consecutively performing the aforementioned rank-2 update twice, with $R = R_1 R_2$, or you can use a direct formula:
	\begin{equation}
		\begin{split}
			R &= \frac{{w}'_l}{{w}_l} = \frac{
			\text{Tr}\left[ e^{-\frac{1}{4} \gamma^T h_{l-1} \gamma} \cdots  e^{-\frac{1}{4} \gamma^T h_{l} \gamma} e^{-\frac{1}{4} \gamma^T m \gamma} \right]
		}{\text{Tr}\left[ e^{-\frac{1}{4} \gamma^T h_{l-1} \gamma} \cdots  e^{-\frac{1}{4} \gamma^T h_{l} \gamma} \right]} \\
		&= \frac{1}{2^N} \eta_m \{ \left[1 - \ima \tanh\left(\lambda\sigma\right) G_{1j, 1k} \right] \left[1 - \ima \tanh\left(\lambda\sigma\right) G_{2j, 2k} \right] + 
		 \tanh(\lambda)^2 \left[ G_{1j, 2j} G_{1k, 2k} - G_{1k, 2j} G_{1j, 2k} \right] \},
		\end{split}
	\end{equation}
	with $\eta_m = (\cosh(\lambda))^2$.
	
	If an auxiliary field flip is accepted based on the acceptance ratio $|R|$, the Green's function can be updated using the Sherman–Morrison–Woodbury identity $\left(I+UCV\right)^{-1}=I-U\left(C^{-1}+VU\right)^{-1}V$. Consider a rank-2 update, let $B_m = e^{-m}$ and $A_l = e^{-h_{l-1}} \cdots e^{-h_{1}} e^{-h_{L_\tau}} \cdots e^{-h_{l}} $. The updated Green's function is then:
	\begin{equation}
		\begin{split}
			G'_l &= 2 \left[ I + A_l B_m \right]^{-1} - I \\
			&= 2 \left[ (I + A_l) + (I+ A_l -I) (B_m - I) \right]^{-1} -I  \\
			&= \left[ I + \frac{I -G_l }{2} (B_m - I) \right]^{-1} (I + G_l) - I \\
			&= \left[ I - U(C^{-1} + VU)^{-1} V \right] (I + G_l) - I \\
			&= G_l - Q^T O Q.
		\end{split}
	\end{equation}
	In this context, $C = e^{-m} - 1 = \begin{pmatrix}
		\cosh(2\lambda) - 1 & -\ima \sigma \sinh(2\lambda) \\
		\ima \sigma \sinh(2\lambda) & \cosh(2\lambda) - 1
	\end{pmatrix}$, $U = \left[ (I - G_l)/2 \right]_{(:), (j;k)}$, and $V$ has all zero elements except for $V_{1, j} = V_{2, k} = 1$. $Q$ is a $2\times 2N$ matrix obtained by extracting the $j$-th and $k$-th rows from the matrix $I+G$, denoted as $Q = (I + G_l)_{(j;k), (:)}$ and $O$ is a $2 \times 2$ matrix with
	\begin{equation}
		O = \begin{pmatrix}
			0 & - \frac{\ima \tanh\left(\lambda\sigma\right)}{1 - \ima \tanh\left(\lambda\sigma\right) G_{j,k} } &  & \\
			+ \frac{\ima \tanh\left(\lambda\sigma\right)}{1 - \ima \tanh\left(\lambda\sigma\right) G_{j,k}} & 0
		\end{pmatrix}.
	\end{equation}
	
	In our simulation of the $t$-$V$ model, we use the method where flipping a single auxiliary field is achieved by performing the rank-2 update twice.
	
	\subsection{E. Numerical Stability}
	
	In PfQMC, as in DQMC, multiplying numerous instances of $e^{-h}$ can lead to numerical instability. Therefore, a similar stabilization scheme, as used in DQMC, is necessary to ensure accurate results.
	
	One advantage of DQMC is that stabilizing the multiplication process automatically stabilizes the sign of the statistical weight. In contrast, in PfQMC, the sign must be evaluated separately through an independent routine. This evaluation can be performed either by directly invoking
	Eq.~\eqref{traceToPfaffian} or by repeatedly applying Eq.~\eqref{eq:productRuleMain}. 
	
	Although the former approach ensures numerically stability, it comes with a computational complexity of $O(N^3 L_{\tau}^3)$. On the other hand, the latter method maintains PfQMC's overall complexity with a cost of $O(N^3 L_{\tau})$, however, it introduces an additional instability. To understand this, consider the scenario depicted in Fig.~\ref{fig:PfQMC1}. 
	As the time step $\Delta \tau \to 0$, multiplying many instances of $e^{-\Delta \tau \frac{1}{4} \gamma^T h \gamma}$ means traversing the zero-weight cracks in small steps. This close proximity to the cracks can cause the Green's function to diverge, resulting in numerical instability during the sign evaluation.
	
	Our current workaround is to track the sign through the fast update procedure. Since the Metropolis sampling guarantees the statistical weight to be non-vanishing, the acceptance ratio $R$ is numerically stable. This allows us to update the sign $s$ using $s_{\text{new}} = \text{sign}\left[ R \right] s_{\text{old}}$. The re-evaluation of this sign only needs to be performed infrequently.
	
	\subsection{F. Lower symmetries and sign-problem-free property}
	
	Sign-problem-free models can be categorized according to the so-called ``lower symmetry condition'' \cite{wei2018semigroup}, a sufficient criterion ensuring that the statistical weight remains positive semi-definite. Intriguingly, all currently known sign-problem-free models are governed by this condition, leading us to wonder if this sufficient condition might also be a necessary one.
	
	Prior to our work, when the sign ambiguity of MQMC remained unresolved, this lower symmetry condition was proved based on the semi-group approach. Now, with the ability to represent this sign as a Pfaffian, it is interesting to re-examine the lower symmetry condition from our new perspective. 
	
	We first briefly review the lower symmetry condition, and the canonical forms of Hamiltonian matrices obeying lower symmetry. If a $2N \times 2N$ matrix $A$ satisfies the following condition:
	\begin{equation}
		U_1^T A U_1 = A^*,\quad \ima(U_2 A - A^* U_2) \leq 0,
	\end{equation}
	with $U_1, U_2$ two anti-commuting real orthogonal matrices, $\{U_1, U_2\} = 0$ and $U_2^2 = -1$, $A^*$ denotes the complex conjugate of $A$, then $A$ is said to obey the lower symmetry. Such matrices are further divided into two classes. The \textit{Majorana class} for $U_1^2 = 1$, and \textit{Kramers class} for $U_1^2 = -1$.

	For the Majorana class, let $P = U_1 U_2$, with $P^2 = 1$ and $\{P, U_1\} = 0$. The $+1$ eigenvectors of $P$ is denoted as $\{\chi_1, \cdots \chi_N\}$, since $U_1$ transforms $+1$ eigenvectors into $-1$ ones. $\{\chi_1, \cdots \chi_N, U_1 \chi_1, \cdots U_1 \chi_N\}$ thus forms a complete basis.
	
	Let $X = (\chi_1, \cdots \chi_N)$ be a $2N \times N$ matrix. In the aforementioned basis, $A$ is transformed into a block matrix $\begin{pmatrix}
		X^T A X & X^T A U_1 X \\
		X^T U_1^T A X & X^T U_1^T A U_1 X
	\end{pmatrix}$, and the off diagonal block satisfy $X^T A U_1 X = X^T A U_2 P X = X^T A U_2 X = -X^T U_2 A^* X = \frac{1}{2} X^T (AU_2 - U_2 A^*) X$. We have
	\eq{
	A = \begin{pmatrix}
		B & \ima C \\
		-\ima C^T & B^*
	\end{pmatrix},
	}{majoranaCanonicalForm}
	with $B$ a skew-symmetric matrix, and $C$ a Hermitian positive semi-definite matrix since $-\ima(U_2 A - A^* U_2)$ is positive semi-definite and $X$ can always be chosen real.
	
	For the Kramers class, $U_1^2 = -1$, we similarly define $Q = U_1 U_2$, here $Q^2 = -1$ and $\{Q, U_1\} = 0$, the complete basis is still $\{\chi_1 \cdots \chi_N, U_1 \chi_1 \cdots U_1 \chi_N \}$ with $\chi_i$ the $+\ima $ eigenvector of $Q$, while $U_1 \chi_i$ the $-\ima$ eigenvector. Physically, $\chi_i$ and $U_1 \chi_i$ corresponds to the complex fermion operators $c_i^\dagger$ and $c_i$.
	
	Now since $Q$ is congruent to a block diagonal matrix $\text{diag}\left[\begin{pmatrix}
		0 & 1 \\
		-1 & 0
	\end{pmatrix}, \cdots \begin{pmatrix}
		0 & 1 \\
		-1 & 0
	\end{pmatrix}\right]$, it is not difficult to show that, $X^*$ and $U_1 X$ are related by a $N\times N$ real orthogonal matrix, denoted by $V$, such that $U_1 X = X^* V$, and $V^2 = -I$. 
	
	Therefore the diagonal matrix can be transformed as
	$X^T A X = -\ima X^T A Q X = X^T U_1 \ima (U_2 A - A^* U_2) X / 2 = V^T X^\dagger R X$, in which $R = -\ima (U_2 A - A^* U_2) / 2 $ is a positive semi-definite matrix, satisfying $U_1^T R U_1 = R^*$. The matrix $\tilde{R} = X^\dagger R X$ also obeys $\tilde{R} \ge 0$, with $V^T \tilde{R} V = X^\dagger U_1^T R U_1 X^* = \tilde{R}^*$. Similarly, $X^T U_1^T A U_1 X = -V^T \tilde{S}$. If summed up, we arrive at the result
	\eq{
	A = \begin{pmatrix}
		V^T R & M \\
		-M^T & -V^T S
	\end{pmatrix},
	}{kramersCanonicalForm}
	with $R, S > 0$, and $V^T M V = M^*, \ V^T R V = R^*, \ V^T S V = S^*$
	
	We might call Eq.~\eqref{majoranaCanonicalForm} and \eqref{kramersCanonicalForm} the canonical forms of Majorana class and of Kramers class, respectively, which is merely a reformulation of the results in \cite{Li2016PRL,wei_majorana_2016,wei2018semigroup}.
	
	It is not difficult to show that whichever lower symmetry $A$ satisfies, so does $\tanh(A/2)$. Following the notation introduced in \cite{wei2018semigroup}, we denote $\eta = \ima U_2$, and the lower symmetry condition can be reformulated as:
	\eq{
	\eta A + A^\dagger \eta \leq 0.
	}{lowerSymDef}
	All matrices $A$ obeying this lower symmetry generate a Lie semi-group, it has been proved that $e^{A^\dagger} \eta e^{A} - \eta < 0$, and we have
	\eq{
	\begin{aligned}
		& \eta \tanh(A/2) + \tanh(A/2)^\dagger \eta < 0 \\ 
		\Leftrightarrow \ &  \eta \left(e^A - I\right) \left(e^A + I\right)^{-1} +  \left(e^{A^\dagger} - I\right) \left(e^{A^\dagger} + I\right)^{-1} \eta < 0 \\
		\Leftrightarrow \ & \left(e^{A^\dagger} + I\right) \eta \left(e^A - I\right) + \left(e^{A^\dagger} - I\right) \eta \left(e^{A} + I\right) < 0 \\
		\Leftrightarrow \ & e^{A^\dagger} \eta e^{A} - \eta < 0,
	\end{aligned}
	}{tanhLowerSym}
	therefore $\tanh(A/2)$ also obeys this lower symmetry condition. 
	
	The sign-problem-free requirement, under our formalism, is nothing but a Pfaffian inequality:
	\begin{equation}
		(-1)^{NL/2} \pfaffian \left( M \right) = (-1)^{NL/2} \pfaffian \begin{pmatrix}
			\tanh \left( \frac{h_1}{2} \right) & -I & I & \cdots \\
			I & \tanh \left( \frac{h_2}{2} \right) &  \ddots & \vdots \\
			 -I  & \ddots & \ddots &  -I \\
			\vdots & \cdots & I & \tanh \left( \frac{h_L}{2} \right)
	\end{pmatrix}  \ge 0.
	\end{equation}
	If every $h_i$'s satisfy the lower symmetry condition, so does the matrix $M$, and $M$ can be also brought into a canonical form as defined by Eq.~\eqref{majoranaCanonicalForm} and \eqref{kramersCanonicalForm}. 
	
	Although \cite{wei2018semigroup} has already proved this inequality, here we show that, for the Majorana class, the Pfaffian-related techniques can offer us a more transparent proof.
	
	It has long been known that Pfaffian can be viewed as a partition function for the dimer model, that is to say, any skew matrix $M$ can be viewed as an adjacent matrix for a specific graph. Each matrix element $M_{ij} \neq 0$ is represented by a link between vertex $i$ and $j$, with weight $|M_{ij}|$. And the Pfaffian is just a sum over all dimer coverings (perfect matching) of such graph \cite{kasteleyn1967graph}. 
	
	As mentioned early, A matrix $M$ belonging to the Majorana class can be written as $M = \begin{pmatrix}
		B & \ima C \\
		-\ima C^T & B^*
	\end{pmatrix}$ with $C$ is a Hermitian positive semi-definite matrix and can be diagonalized as $U^\dagger C U = \Lambda = \text{diag}\left( \lambda_1, \cdots \lambda_{NL} \right)$, $\lambda_i > 0$. we have
	\begin{equation}
		\pfaffian\left( M \right) = \pfaffian\left( Q^T M Q \right) = \pfaffian\begin{pmatrix}
			B' & \ima \Lambda \\
			-\ima \Lambda & B'^*
		\end{pmatrix}, \, \text{with } Q = \begin{pmatrix}
			U & 0 \\
			0 & U^*
		\end{pmatrix} \text{ and } B' = U^T B U.
	\end{equation}
	The transformed $M' = \begin{pmatrix}
		B' & \ima \Lambda \\
		-\ima \Lambda & B'^*
	\end{pmatrix}$ can thus be represented by a graph:
	
	\begin{figure}[h]
		\centering
		\begin{tikzpicture}  
			\draw[fill=gray!50, draw=gray!50] (0,0) ellipse (3cm and 0.5cm);  
			  
			\node[fill=black,circle,inner sep=1pt, label=left:$i$] (A) at (-1,0.3) {};  
			\node[fill=black,circle,inner sep=1pt] (B) at (1,-0.3) {};  
			\node[fill=black,circle,inner sep=1pt] (C) at (-0.5,-0.35) {};
			\node[fill=black,circle,inner sep=1pt, label=right:$j$] (D) at (1.3,+0.35) {};
			  
			\draw (A) -- (B);  
			\draw (B) -- (C);  
			\draw (C) -- (A);  
			\draw (D) -- node[midway, above, sloped] {$B'_{ij}$} (A);  
			\draw (C) -- (D);  
	
			\draw[fill=gray!50, draw=gray!50] (0,-2) ellipse (3cm and 0.5cm);  
			  
			\node[fill=black,circle,inner sep=1pt] (A1) at (-1,0.3-2) {};  
			\node[fill=black,circle,inner sep=1pt] (B1) at (1,-0.3-2) {};  
			\node[fill=black,circle,inner sep=1pt] (C1) at (-0.5,-0.35-2) {};
			\node[fill=black,circle,inner sep=1pt] (D1) at (1.3,+0.35-2) {};
			  
			\draw (A1) -- (B1);  
			\draw (B1) -- (C1);  
			\draw (C1) -- (A1);  
			\draw (D1) -- node[midway, above, sloped] {$B'^*_{ij}$} (A1);  
			\draw (C1) -- (D1);  
	
			\draw[blue!80] (A) -- node[midway, left] {$\ima \lambda_{i}$} (A1);
			\draw[blue!80] (D) -- node[midway, right] {$\ima \lambda_{j}$} (D1);
			\draw[blue!80] (B) -- (B1);
			\draw[blue!80] (C) -- (C1);
		\end{tikzpicture}  
	\end{figure}
	Here the matrices $B'$ and $B'^*$ can be drawn as two identical graphs in the upper and lower plane, respectively. Each link in the lower plane bears a weight that is the complex conjugate of its corresponding link in the upper plane.
	The $(-1)^{N L/2}$ prefactor can be absorbed by a permutation of the vertices, such that each vertex in the ``upper plane" always has smaller index than those in the "lower plane". Also, from the knowledge of perfect matching, we can easily find that $\ima \lambda $ must be presented in the final expression in pairs, with each pair contributing both an $\ima^2=-1$, and an additional $(-1)$ because of permutation effects. Therefore
	\begin{equation}
		(-1)^{NL/2} \pfaffian\left( M \right) = a_0 + \sum_{ij} a_{1, ij} \lambda_i \lambda_j + \sum_{ijkl} a_{2, ijkl} \lambda_i \lambda_j \lambda_k \lambda_l + \cdots  \geq 0,
	\end{equation}
	with each $a$ a positive coefficients. The Pfaffian inequality is easily proved in this way.

	\subsection{G. Kitaev Chain with nearest-neighbor interaction $V$}
	
	\begin{figure*}[t!]
		\includegraphics*[width=0.5\linewidth]{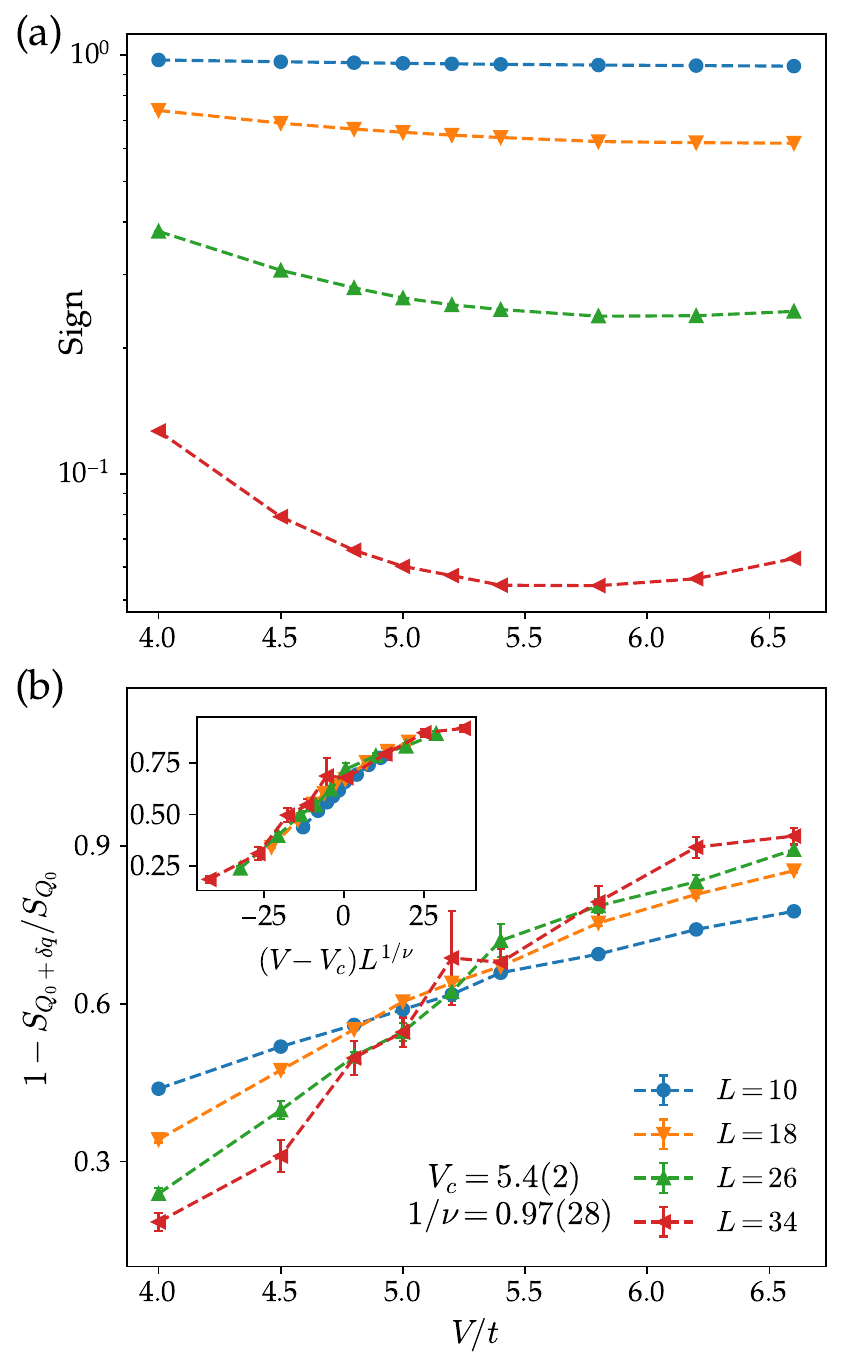}
		\caption{The Kitaev chain at $\Delta = t$, with chemical potential $\mu = 1.0$. The number of imaginary time slices is fixed as $L_\tau = L$, $\Delta \tau=0.1$. (a) The average sign. Upon doping, the sign problem becomes worse. (b) The RG-invariant ratio $R = 1 - S_{Q_0 + \delta q} / S_{Q_0}$, same as in Fig.\ref{fig:chain-sign-z2}(e). Here it captures the quantum phase transition between the ICDW phase and the CDW phase, with critical interaction estimated to be $V_c = 5.4(2)$, in accordance with \cite{miao2018majorana}.}
		\label{fig:chain-dope}
	\end{figure*}
	
	Special efforts have been paid to the Kitaev chain model with $\Delta = t$ and $\mu = 0$. Such an interacting Kitaev chain can be written as:
	\begin{equation}
		\begin{split}
			H &= \sum_{j=1}^{L-1} [ (-t c^\dagger_{j} c_{j+1} + \Delta c^{\dagger}_{j+1} c^\dagger_{j} + \text{h.c.} ) +  V (n_{j} - 1/2)(n_{j+1} - 1/2) ] \\
			&= \sum_{j=1}^{L-1} \left[ \frac{\ima}{2} \left[(-1)^j t + \Delta\right] \gamma^1_{j} \gamma^1_{j+1} + \frac{\ima}{2}\left[(-1)^j t - \Delta\right] \gamma^2_{j} \gamma^2_{j+1} \right] - \frac{V}{4} \sum_j \gamma_j^1 \gamma_j^2 \gamma_{j+1}^1 \gamma_{j+1}^2.
		\end{split}
	\end{equation}
	This is related to the form provided in \cite{miao2017exact} through the transformation $\gamma^1 \to \gamma^2, \, \gamma^2 \to -\gamma^1$ for odd $j$'s.
	
	And if the system is away from half-filling, an additional chemical potential term is given by:
	\begin{equation}
			H_\mu = -\mu \sum_{j = 1}^{L-1} \left(c^\dagger_j c_j - \frac{1}{2}\right)  =  -\frac{\ima \mu} {2} \sum_{j} \gamma^1_j \gamma^2_j .
	\end{equation}
	This model exhibits a $Z_2$ symmetry that comes from the fermion parity. If it is inside the topological superconducting (TSC) phase, the two-fold degenerate ground states can be distinguished by this $Z_2$ symmetry, and one would like to only explore the invariant subspaces of this symmetry. The $Z_2$ charge can be defined as $Z^f_2 = \prod_{i=1}^N (2 c^\dagger_i c_i - 1) = (-\ima)^{N} \gamma_1^1 \gamma_1^2 \cdots \gamma_N^1 \gamma_N^2$, and a projection $P$ onto the $+1$ ($-1$) subspace of this $Z^f_2$ can be defined as ${\hat P}_{\pm} = (1 \pm Z^f_2) / 2$. The PfQMC algorithm needs little modification to handle this symmetry constraint, as can be seen from:
	\begin{equation}
		\begin{split}
			\braket{\hat O}_{Z_2^f = \pm 1} &= \frac{\trace{{\hat P}_{\pm} \hat{O} e^{-\beta H} {\hat P}_{\pm}}}{\trace{{\hat P}_{\pm} e^{-\beta H} {\hat P}_{\pm}}} = \frac{\sum_{\{\phi_n\}} \trace{{\hat P}_{\pm} \hat{O} e^{-h(\{\phi_n\})}}}{\sum_{\{\phi_n\}} \trace{{\hat P}_{\pm} e^{-h(\{\phi_n\})}} } \\
			&= \frac{\sum_{\{\phi_n\}} {w}(\{ \phi_n\}) \braket{{\hat P}_{\pm} \hat O}_{\{\phi_n\}}}{\sum_{\{\phi_n\}} {w}(\{ \phi_n\}) \braket{{\hat P}_{\pm}}_{\{\phi_n\}}} = \frac{ \Braket{ \text{sign}(\{ \phi_n\}) \braket{\hat P_{\pm}\hat O}_{\{ \phi_n\}} }_{\{\phi_n\}  \sim |{w}| } }{ \Braket{ \text{sign}(\{ \phi_n\}) \braket{\hat P_{\pm}}_{\{ \phi_n\}} }_{\{\phi_n\}  \sim |{w}| }  }. \\
		\end{split}
	\end{equation}
	in which the expectation value of a product of Majorana operators is given by:
	\begin{equation}
		\Braket{\gamma_{i_1} \cdots \gamma_{i_{2s}}} = \pfaffian_{1\leq k < l \leq 2s} \Braket{\gamma_k \gamma_l}
	\end{equation}
	as a direct consequence of the Wick's theorem \cite{balian_nonunitary_1969}. This allows us to evaluate the finite temperature expectation value of any observable $\hat O$, in a symmetry sector defined by the projection operator ${\hat P}_{\pm}$.
	
	Upon doping, the model exhibits a sequence of quantum phase transitions from TSC to incommensurate CDW (ICDW) phase, and to the usual CDW phase with $k = \pi$, as already studied in \cite{miao2018majorana}. Note that if one focus on the quantum phase transition from CDW to the ICDW phase, the ICDW phase breaks the lattice translational symmetry, hence is expected to be of quasi-long-range-order (QLRO), as described by the Berezinskii-Kosterlitz-Thouless (BKT) theory.
	
	In our numerical study, the sign problem generally worsens with increasing $\mu$, making it difficult to detect quasi-long-range order (QLRO). Therefore, we focus on the charge-density-wave (CDW) order parameter, employing the renormalization group invariant ratio $R = 1 - S_{Q_0 + \delta q} / S_{Q_0}$, as already discussed in the main article, where $S_{Q}$ is the CDW structure factor defined as $\frac{1}{L^2} \sum_{i, j} e^{\ima Q (i-j)} \braket{(n_i - 1/2) (n_j - 1/2)}$, $Q = \pi$ and $\delta q = 2\pi / L$. As an illustrative example, we fix $\mu = 1.0$, the average sign and the ratio $R$ are depicted in Fig.~\ref{fig:chain-dope}. Our PfQMC algorithm remains valid away from the exact-solvable point, aligning with our expectations.
	
	\subsection{H. Majorana resolved rign of the spinless $t$-$V$ Model on a honeycomb lattice}
	\begin{figure*}[t!]
		\includegraphics*[width=0.8\linewidth]{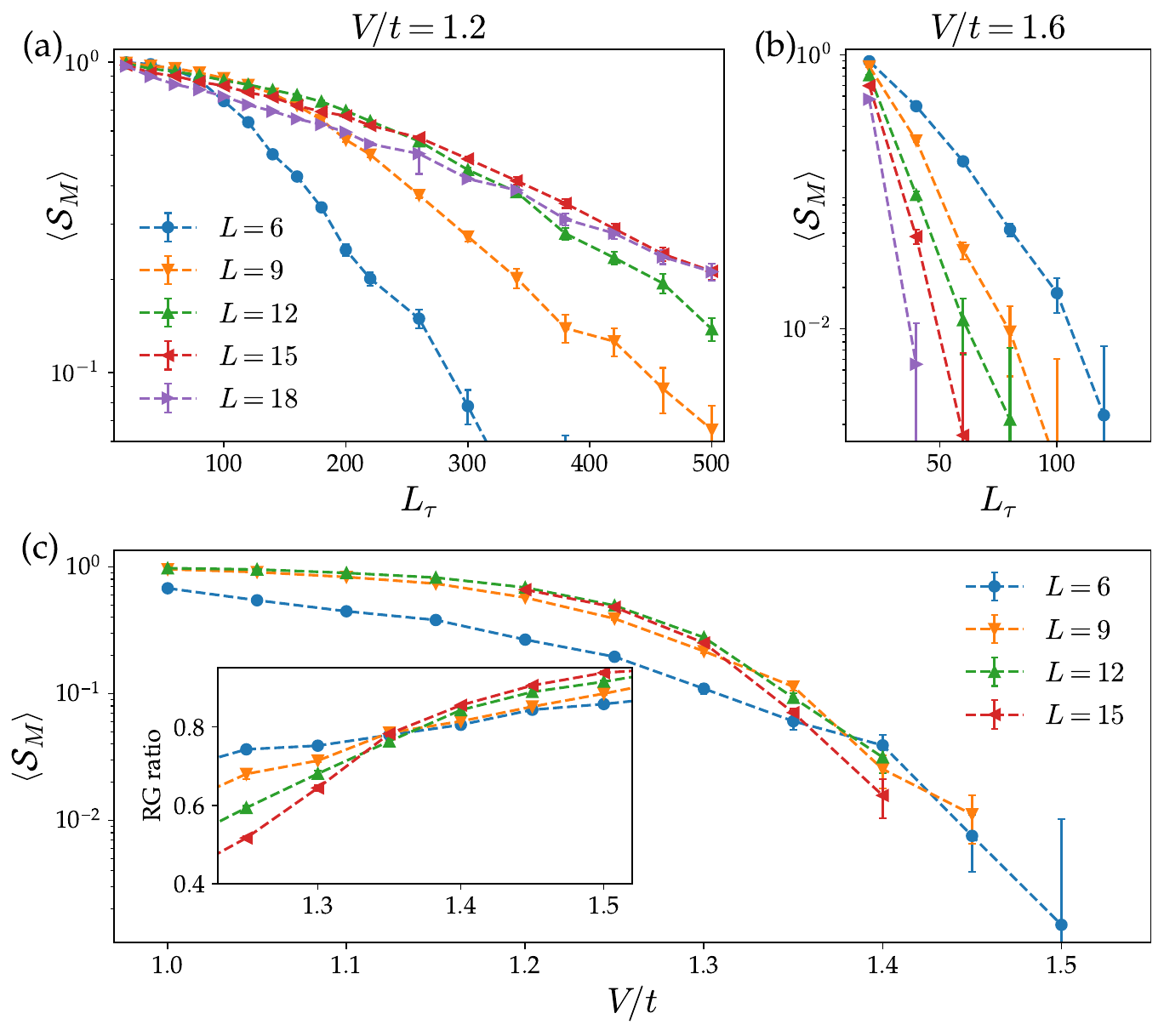}
		\caption{Dependence of the Majorana resolved sign $\braket{\mathcal{S}_M}$ on the lattice size $L$ and the inverse temperature $\beta = \Delta \tau L_\tau$, with $\Delta\tau = 0.1/t$. (a) For $V < V_c \approx 1.36t$ \cite{wang2016stochastic}, at low enough temperature (large enough $L_\tau$), the average sign grows with increasing lattice size $L$. (b) For $V > V_c$, the average sign reduces at any temperature. (c) For fixed $L_\tau = 200$, $\braket{\mathcal{S}_M}$ vs interaction strength $V$. Inset demonstrates the RG-invariant ratio $R = 1 - S_{\bm Q_0 + \delta \bm q} / S_{\bm Q_0}$ defined accordingly for the CDW phase, through which a phase transition at $V_c \approx 1.35$ is observed, in accordance with the previous studies \cite{li2015fermion,wang2016stochastic}. Around $V_c$, the resolved sign roughly shows a ``crossing'' behavior.}
		\label{fig:mrsign}
	\end{figure*}
	
	The spinless $t$-$V$ model on a honeycomb lattice has been extensively studied \cite{wang2014fermionic,wang2016stochastic,li2015fermion}, and through the use of an HS transformation defined in Eq.~\eqref{eq:HSforSymMajorana}, the partition function can be written as \cite{Li2015PRB}
	\begin{equation}
		Z = \sum_{\{\phi_n\}} w^2(\{\phi_n\}),
	\end{equation}
	with $w(\{\phi_n\})$ the ``weight'' associated with one individual Majorana flavour, and is guaranteed to be a real number. The model is sign-problem-free since $W(\{\phi_n\}) = w^2(\{\phi_n\}) \ge 0$. Motivated by previous works \cite{mondaini2022quantum,mondaini2023universality} where the average sign of a single spin species is studied, the average sign of a single Majorana flavour can be similarly defined here. We name it the Majorana-resolved sign:
	\begin{equation}
		\braket{\mathcal{S}_M} = \sum_{\{\phi_n\}} \text{sgn}\left[ w(\{\phi_n\}) \right] W(\{\phi_n\}) / \sum_{\{\phi_n\}} W(\{\phi_n\}).
	\end{equation}
	
	It was argued in \cite{mondaini2023universality} that the average sign of a single species contains information about the quantum criticality. Here we observe a similar behavior, as depicted in Fig.~\ref{fig:mrsign}. We pick two interaction strength $V=1.2t < V_c$ ($V=1.6t > V_c$) to represent the semi-metal (CDW) phase. In the CDW phase, the average sign exhibits a usual exponentially decaying feature, but in the semi-metal phase, average sign in larger system sizes tends to decay much slower with respect to increasing $L_\tau$ than in smaller sizes. At sufficiently low temperatures, the average sign increases with increasing $L$. 
	This low-temperature behavior is similar to what has been observed in \cite{mondaini2023universality}. But there are two major differences: (1) the average sign in \cite{mondaini2023universality} grows (with respect to $L$) unanimously at any temperature, but here the growth happens only when $L$ is smaller than some critical $L_c$, and $L_c$ also grows while lowering the temperature. (2) In \cite{mondaini2023universality}, the average sign can be used as an observable to extract the quantum critical behaviors, this relies on the fact that the resolved sign at quantum critical point $V_c$ is saturated at some finite (and moderately large) value. But in our case, this average sign becomes too small to do any in-depth quantitative analysis, as displayed in Fig.~\ref{fig:mrsign}(c). The origin of this non-trivial behavior awaits to be further explored.
	
	\end{widetext}

\end{document}